\documentclass[aip,reprint]{revtex4-1}
\DeclareUnicodeCharacter{0301}{\'e}
\usepackage{graphicx}
\usepackage{dcolumn}
\usepackage{bm}

\usepackage[utf8]{inputenc}
\usepackage[T1]{fontenc}
\usepackage{mathptmx}
\usepackage{etoolbox}
\usepackage{xcolor}
\usepackage{placeins} 
\usepackage{amsmath} 
\usepackage{makecell}
\usepackage{float}
\usepackage{setspace} 

\makeatletter
\def\@email#1#2{%
 \endgroup
 \patchcmd{\titleblock@produce}
  {\frontmatter@RRAPformat}
  {\frontmatter@RRAPformat{\produce@RRAP{*#1\href{mailto:#2}{#2}}}\frontmatter@RRAPformat}
  {}{}
}%
\makeatother
\begin{document}

\preprint{AIP/123-QED}

\title{Sub-cavity Induced Passive Control of Confined Supersonic Cavity Flows Across Varying Freestream Mach Numbers}
\author{Sreejita Bhaduri}\affiliation{%
Department of Aerospace Engineering, Indian Institute of Technology Kanpur, 208016, Kanpur, India
}

\author{Mohammed Ibrahim Sugarno}
 \affiliation{%
Department of Aerospace Engineering, Indian Institute of Technology Kanpur, 208016, Kanpur, India
}
\author{Ashoke De}%
 \email{ashoke@iitk.ac.in.}
 \affiliation{%
Department of Aerospace Engineering, Indian Institute of Technology Kanpur, 208016, Kanpur, India
}

\date{\today}

\begin{abstract}  
The self-sustaining cavity oscillations enhance fluid mixing, promoting energy and momentum transport. Despite their utility, the associated oscillation frequencies can amplify acoustic loading, potentially damaging structures that house cavities. Understanding cavity flow dynamics across various geometries and freestream conditions is, therefore, crucial, along with devising strategies to regulate these oscillations without compromising performance. In this study, we investigate the role of sub-cavities placed at the front and aft walls of a cavity confined by a top wall with a deflection angle of $2.29^\circ$, under freestream Mach numbers $M_\infty = 2$ and $3$. We perform large-eddy simulations (LES) using OpenFOAM for each configuration and analyze the unsteady pressure signals through spectral methods. The results show that the aft-wall sub-cavity most effectively suppresses the dominant oscillation frequency at the lower Mach number ($M_\infty = 2$), while the front-wall sub-cavity achieves the greatest suppression at the higher Mach number ($M_\infty = 3$).We examine the density gradient contours (numerical Schlieren), vorticity fields, normalized acoustic impedance, and global wavelet power to identify the flow mechanisms responsible for frequency attenuation. At $M_\infty = 2$, the aft-wall sub-cavity entrains mass and interrupts the convectively driven feedback loop. At $M_\infty = 3$, the front-wall sub-cavity weakens the hydrodynamic–acoustic coupling between the cavity pressure waves and the shear layer near the leading edge, thereby disrupting the compressibility-driven feedback loop. We observe that these configurations suppress the major oscillation frequencies by 5.45\%  and 23.4\%  for $M_\infty = 2$ and $3$, respectively. We further validate these findings by computing the cross-correlation between pressure probes located near the front and aft edges and by analyzing the spatio-temporal coherence in pressure and velocity fields using Dynamic Mode Decomposition (DMD), which confirms the mechanisms behind the observed frequency suppression.
\end{abstract}

\maketitle


\begin{figure*}

	\includegraphics[scale=0.425]{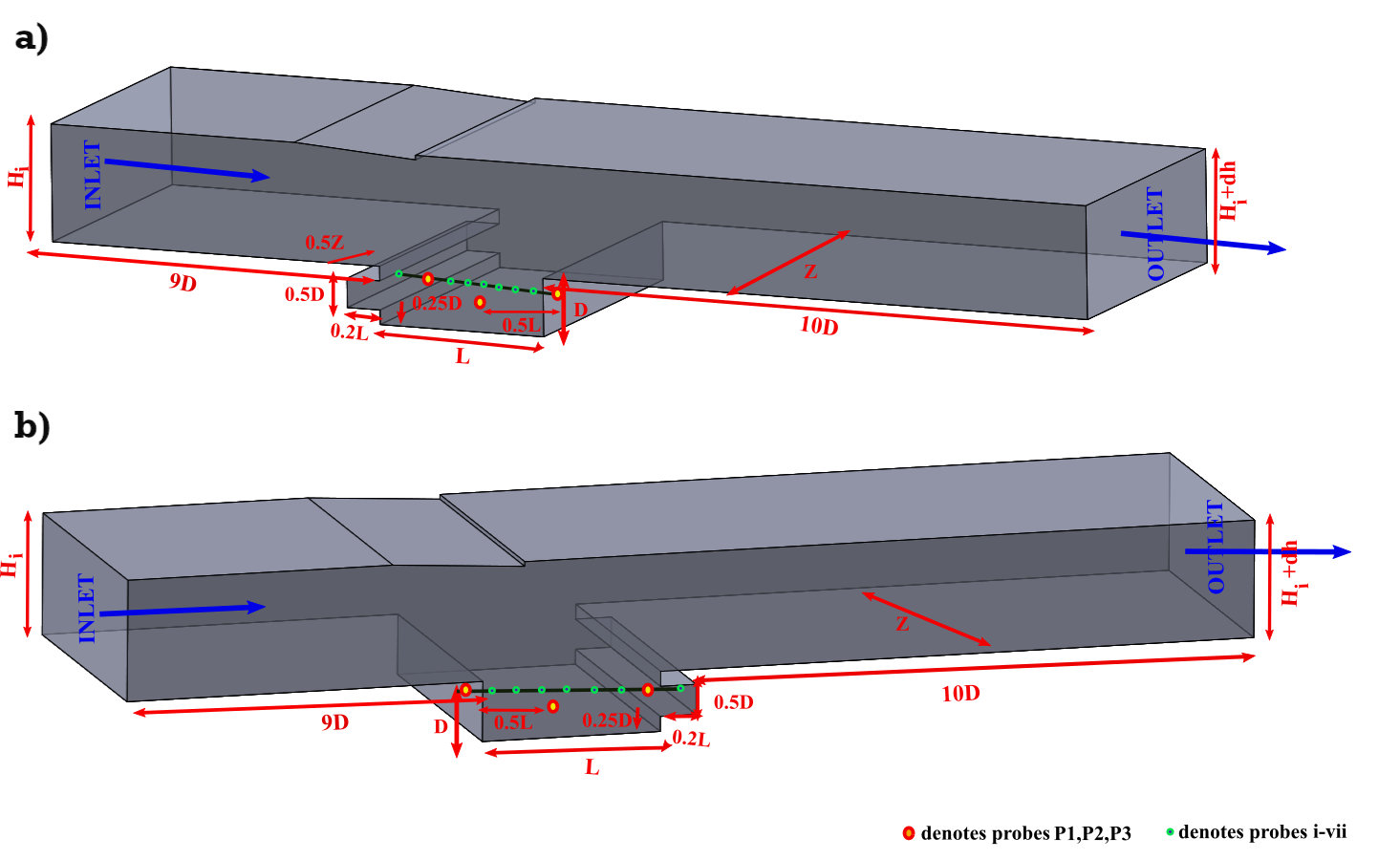}
    \centering

    \caption{\label{fig:1} Schematic of the confined cavity with a) front-wall b) aft-wall subcavity with probe locations.}
   \end{figure*}
  \begin{figure}[htbp]
    \centering
    \includegraphics[scale=0.27]{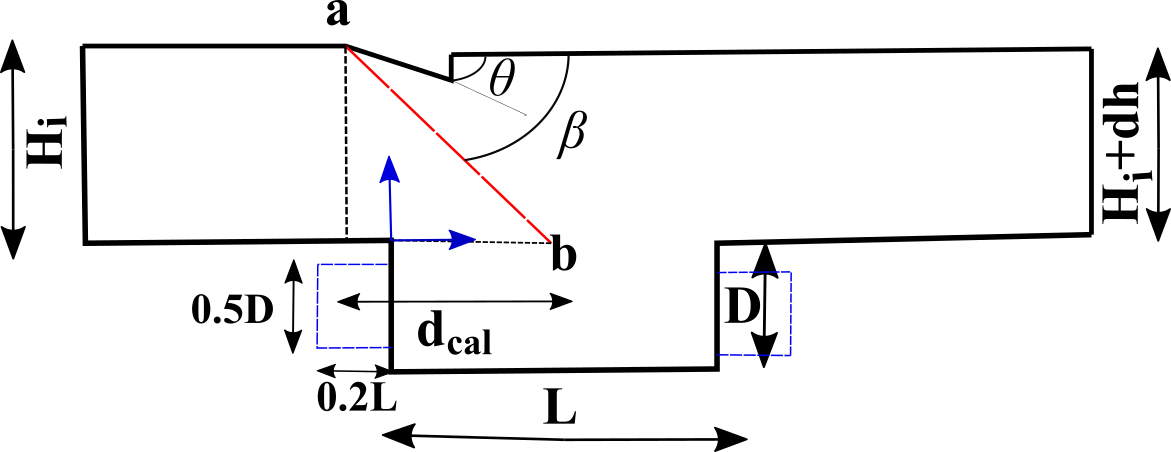}
    \caption{\label{fig:2} Schematic of cavity geometry illustrating the deflection angle ($\theta$), the shock angle ($\beta$), the origin of the shock (a), and the impinging point (b).}
\end{figure}

    \begin{figure}

	\includegraphics[scale=0.425]{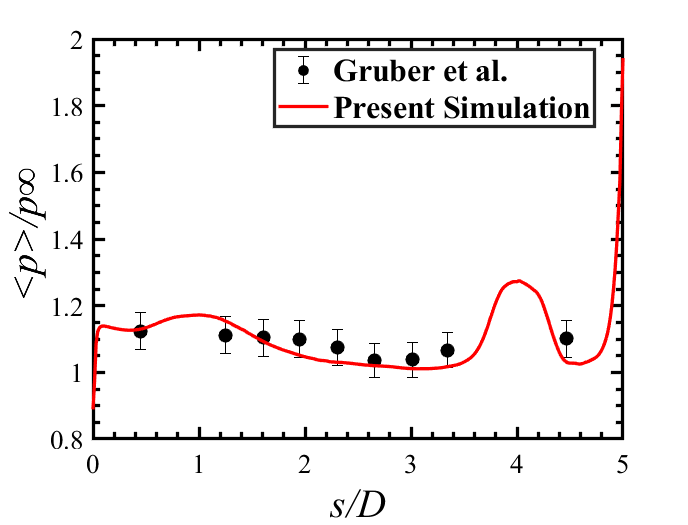}
    \centering

    \caption{\label{fig:3} Validation of the present simulation against the experimental data of Gruber et al.\cite{gruber2001fundamental}.}
   \end{figure}

\section{\label{sec:level1}Introduction}
Cavities with self-sustaining, feedback-driven oscillations improve fluid mixing, which in turn enhances momentum and energy exchange. These oscillations are essential for several industrial applications, such as heat exchangers, weapon bays, air-fuel mixing in scramjet combustors, and promoting nuclear reactor coolant flow~\cite{charwat1961investigation,emery1969recompression,saravanan2020isolator,sitaraman2021adaptive,johnson2010instability,chakravarthy2018analytical,devaraj2020experimental,sekar2020unsteady}.
The cavity’s aspect ratio, defined as the ratio of its length (L) to depth (D), strongly influences the flow field, along with external factors such as Mach number, incoming boundary layer thickness, and top wall confinement~\cite{plentovich1993experimental,tracy1992measurements}.
Researchers classify cavities based on their length-to-depth ratio ($L/D$), identifying open cavities when $L/D < 10$ and closed cavities when $L/D > 13$\cite{stallings1987experimental}. Closed cavities, which exhibit higher drag coefficients, serve well in heat exchange applications\cite{wang2013characteristics,heller1971flow} but lack significant acoustic behavior. 

In contrast, open cavities commonly found in aerospace systems depend on a balance between freestream energy input and dissipation mechanisms such as acoustic radiation, viscous losses, and convective mass exchange. In these cavities, the shear layer separates from the leading edge, grows across the cavity, and impinges on the aft wall~\cite{heller1971flow,heller1975physical,heller1996cavity,krishnamurty1955acoustic}. This interaction generates a reattachment shock at the trailing edge and promotes alternating inflow and outflow of mass within the cavity. As the fluid enters the cavity, it produces a pressure wave that travels upstream, reflects off the front wall, and interacts with the separating shear layer. This reflection amplifies disturbances through acoustic–vortex coupling, which then grow and once again impinge on the aft wall—forming a self-sustaining feedback loop.

Open cavities typically exhibit two distinct acoustic features: (a) discrete tones, known as Rossiter modes, generated by interactions such as shear–wall, shock–shear layer, vortex–vortex, vortex–wall, or vortex–shear layer coupling; and (b) broadband noise, which arises from turbulent fluctuations, shear layer dynamics, and interactions with the freestream~\cite{lawson2011review}.

Given the structural risks associated with these oscillations, researchers have prioritized the development of effective flow control strategies to regulate or suppress the dominant tones~\cite{rockwell1978self,rowley2006dynamics}. Rowley categorizes control techniques into two types~\cite{rowley2006dynamics}: active control, which uses actuators or external energy sources to dynamically manipulate the flow, and passive control, which employs fixed geometric modifications or flow obstructions to alter the flow field without requiring additional energy input.

Zhuang et al.~\cite{zhuang2006supersonic} reduced tone amplitudes by 20 dB and total noise by 9 dB by using continuous and pulsed mass injection inside the cavity. According to Yilmaz et al.~\cite{yilmaz2013numerical} laser energy deposition within the cavity can reduce noise levels by as much as 7 dB. Vikramaditya and Kurian~\cite{vikramaditya2009pressure} investigated the impact of ramp angle deflection and discovered that pressure oscillations were significantly reduced when the ramp angle was decreased from $90^\circ$ to $60^\circ$.
Numerous studies have highlighted how compression waves, which are produced at the trailing edge of the cavity, regulate mass exchange and the production of acoustic waves. Malhotra and Vaidyanathan ~\cite{malhotra2016aft} reduced sound intensity by introducing an aft-wall offset. According to Lee et al. ~\cite{lee2008passive} sub-cavities at the trailing edge reduce noise more efficiently than blowing techniques or surface roughness. Alam et al. ~\cite{alam2007new} reported that sub-cavities substantially dampened pressure oscillations in open supersonic cavities, with suppression effectiveness depending on the sub-cavity length. Lad et al.~\cite{lad2018experimental} demonstrated that the sub-cavities are straightforward yet efficient passive control techniques. He altered the sub-cavity length that induced a mode switch from fluid-dynamic to fluid-resonant oscillations. Panigrahi et al. \cite{panigrahi2019effects} investigated experimentally and numerically the effect of sub-cavity placement on flow oscillations in a rectangular open cavity at Mach 1.71. It demonstrates how the sub-cavity’s size and position have a significant impact on the kind and intensity of oscillations. The aft-wall sub-cavity functions as a resonator, changing modal frequencies, whereas the front-wall sub-cavity reduces the dominant tone by 34.1 dB. The sub-cavities produce the most significant suppression when positioned at both walls, lowering the overall sound pressure by 14.5 dB and the dominant tone by 34.9 dB.
Jain and Vaidyanathan ~\cite{jain2021aero} conducted numerical simulations on a baseline cavity with a length-to-depth (L/D) ratio of 2 at Mach numbers 1.71 and 3.25. Since the cavity lacked top wall confinement, compression waves did not impinge upon the shear layer. They investigated a range of sub-cavity lengths at the front and rear walls, detecting unique oscillation patterns and fluid-resonant modes. They used flow visualization techniques in their investigation, such as numerical Schlieren based on streamlines, vorticity contours, and density gradients. Their findings demonstrated that front-wall sub-cavities successfully suppress self-sustained oscillations. Chavan et al.~\cite{chavan2022experimental} experimentally investigated a cavity with an L/D ratio of 2 and freestream Mach number of 1.71, incorporating sub-cavities of length ratio 0.2 at both the front and aft-walls. Their findings corroborated that front-wall sub-cavities play a key role in mitigating cavity oscillations. Jain et al.~\cite{jain2023effects} conducted trials on an unconfined cavity with a similar structure in a follow-up study, adding floor sub-cavities in addition to front and aft-wall sub-cavities. They concluded that whereas front-wall sub-cavities effectively suppress cavity oscillations, floor and aft-wall sub-cavities act as passive resonators.

\begin{table}[ht]
    \centering
    \caption{\label{tab:table1}Summary of the cavity configurations under study along with the names by which they are referred to in the article.}
    \begin{ruledtabular}
    \begin{tabular}{cc}
    \textbf{Case} & \textbf{Position of Subcavities}  \\
         \hline
      \mbox{Reference case (R)} & \mbox{-} \\  
      \mbox{Front-wall (F)} & \mbox{  midpoint of the front-wall } \\  
       \mbox{Aft-wall (A)} & \mbox{ midpoint of the aft-wall } \\ 
    \end{tabular}
    \end{ruledtabular}
\end{table}

\begin{table}[ht]
    \centering
    \caption{\label{tab:table2} Coordinates of the internal probes and probes i-vii according to the cavity configuration. All the coordinates are expressed in terms of the length (L) and depth (D) of the cavity.}
   \begin{ruledtabular}
    \begin{tabular}{cc}
      \multicolumn{2}{c}{\textbf{Reference case (R)}} \\
      \hline
        \textbf{Probe} & \textbf{Coordinates of probes}  \\
         \hline
      \mbox{i} & \mbox{(0.125L,-0.5D,2.5D)} \\  
      \mbox{ii} & \mbox{ (0.25L,-0.5D,2.5D)} \\  
       \mbox{iii} & \mbox{ (0.375L,-0.5D,2.5D) } \\ 
       \mbox{iv} & \makecell{ (0.5L,-0.5D,2.5D)} \\ 
       \mbox{v} & \makecell{ (0.625L,-0.5D,2.5D)} \\
       \mbox{vi} & \makecell{ (0.75L,-0.5D,2.5D)} \\
       \mbox{vii} & \makecell{ (0.875L,-0.5D,2.5D) } \\
       \hline
        \multicolumn{2}{c}{\textbf{Front-wall sub-cavity case (F)}} \\
        \hline
        \textbf{Probe} & \textbf{Coordinates of probes}  \\
         \hline
      \mbox{i} & \mbox{(-0.05L,-0.5D,2.5D)} \\  
      \mbox{ii} & \mbox{  (0.1L,-0.5D,2.5D) } \\  
       \mbox{iii} & \mbox{ (0.25L,-0.5D,2.5D) } \\ 
       \mbox{iv} & \makecell{ (0.4L,-0.5D,2.5D)} \\ 
       \mbox{v} & \makecell{ (0.55L,-0.5D,2.5D)} \\
       \mbox{vi} & \makecell{(0.7L,-0.5D,2.5D)} \\
       \mbox{vii} & \makecell{ (0.85L,-0.5D,2.5D)} \\
       \hline
        \multicolumn{2}{c}{\textbf{Aft-wall sub-cavity case (A)}} \\
        \hline
        \textbf{Probe} & \textbf{Coordinates of probes}  \\
         \hline
      \mbox{i} & \mbox{(0.15L,-0.5D,2.5D)} \\  
      \mbox{ii} & \mbox{  (0.3L,-0.5D,2.5D) } \\  
       \mbox{iii} & \mbox{ (0.45L,-0.5D,2.5D) } \\ 
       \mbox{iv} & \makecell{ (0.6L,-0.5D,2.5D)} \\ 
       \mbox{v} & \makecell{ (0.75L,-0.5D,2.5D)} \\
       \mbox{vi} & \makecell{ (0.9L,-0.5D,2.5D)} \\
       \mbox{vii} & \makecell{ (1.05L,-0.5D,2.5D) } \\
    
   \end{tabular}
    \end{ruledtabular}
\end{table}

Previous studies on passive control of cavity oscillations have primarily focused on open cavities without confinement. However, top wall confinement, a common feature in practical supersonic applications such as scramjet combustors, introduces compression waves that significantly alter the cavity flow dynamics~\cite{karthick2021shock}. In our earlier work~\cite{bhaduri2024effects}, we performed three-dimensional simulations of confined cavities and showed that impinging shocks interact with shear layer disturbances and reflect as expansion fans, which intensify Kelvin–Helmholtz (KH) roll formation. The resulting compression–expansion cycles strengthen KH instability and increase energy transfer from the mean flow to fluctuations. Consequently, confined cavities with impinging shocks exhibit lower oscillation frequencies than cavities without such shock interactions. The shock impingement location on the shear layer plays a critical role in this frequency reduction~\cite{bhaduri2025influence}.

In a recent study~\cite{bhaduri2024flow}, we investigated the effect of sub-cavity length on oscillation suppression for confined cavity flows at a freestream Mach number of \(M_\infty = 1.71\). Sub-cavities were placed at the midpoints of the front and aft walls of a cavity with aspect ratio \(L/D = 3\), considering length ratios of \(l/L = 0.2\), 0.25, and 0.3. The results showed that the variation of oscillation suppression with sub-cavity length is non-monotonic. Two-dimensional RANS simulations revealed that a front-wall sub-cavity of \(l/L = 0.2\) most effectively suppressed the dominant oscillation frequency, achieving a reduction of nearly 60\%. This suppression resulted from the dissipation of pressure waves interacting with vortices at the sub-cavity edge, disrupting the feedback loop without significantly perturbing the separating shear layer. However, increasing the sub-cavity length beyond \(l/L = 0.2\) led to a re-establishment of the feedback mechanism, resulting in higher dominant frequencies and stronger shear layer disturbances. For aft-wall sub-cavities, particularly at \(l/L = 0.2\), the dominant frequency dropped by approximately 7\%, as the upstream-traveling pressure waves were attenuated before reaching the leading edge Based on these observations, the l/L =0.2 configuration represents a practical balance between geometric simplicity and flow control effectiveness.

These two-dimensional simulations, however, did not capture the spanwise evolution of Kelvin–Helmholtz vortices—an essential feature of three-dimensional cavity flows. To address this limitation, the present study extends the investigation to three-dimensional cases at higher Mach numbers, incorporating both front-wall and aft-wall sub-cavities. The objectives are to:
\begin{itemize}
    \item Assess the effectiveness of a sub-cavity with \(l/L = 0.2\) in suppressing oscillation frequencies in a confined supersonic cavity (\(L/D = 3\)) influenced by an impinging shock.
    \item Examine how frequency attenuation varies at \(M_\infty = 2\) and \(M_\infty = 3\), depending on sub-cavity placement at the front or aft wall.
    \item Identify the mechanisms responsible for frequency suppression and analyze the flow field modifications induced by different sub-cavity configurations.
\end{itemize}

The article is structured as follows: Section \ref{sec:level2} covers the geometrical configuration, numerical methods, boundary conditions, study of grid independence, and validation against experiments. Section \ref{sec:level3} details the results and their physical interpretation, discussing the variation in the dominant frequency in the different cavity configurations and the flow field at different Mach numbers due to the incorporation of the sub-cavities. Section \ref{sec:level4} summarizes the crucial findings of this numerical investigation.

\section{\label{sec:level2}Numerical Methodology}

\subsection{Geometrical Configuration}
This study analyzes confined cavities with a top wall deflection angle of $2.29^\circ$. We compare the flow fields and spectral characteristics between a reference cavity (without sub-cavities) and two modified configurations that include a sub-cavity of 0.2L, positioned at the midpoint of (a) the front wall and (b) the aft wall. Fig.~\ref{fig:1} presents the schematic of these confined cavity configurations with sub-cavities

To ensure the development of a fully turbulent inflow, we place the cavity at a distance of 9D downstream of the inlet. To maintain smooth flow throughout the domain, we extend the outlet 10D downstream of the cavity's trailing edge. These inlet and outlet distances remain consistent across all configurations studied.analysesThe intake and outlet heights are set to $H_i=2.021D$ and $H_o=1.954D$, respectively, and the resulting height difference ($\Delta h$) preserves sufficient mass flow rate despite the pressure loss caused by the oblique shock at the deflection corner.
To capture three-dimensional effects, we extend the computational domain 5D in the spanwise direction (Z = 5D). Table~\ref{tab:table1} summarizes all cavity configurations analyzed in this study.

We place several pressure probes along the mid-span plane (Z=2.5D) to monitor the flow dynamics within the cavity. As indicated by red dots in Fig. \ref{fig:1}, probe P3 (0.5L, -0.5D, 2.5D) is positioned at the midpoint of the cavity floor, while the probes P1 (0, -0.5D, 2.5D) and P2 (L, -0.5D, 2.5D) are situated close to the cavity's leading and trailing edges, respectively. In the reference configuration, P1 and P2 align with the geometric centers of the front and aft walls. Additionally, we deploy seven equally spaced internal probes along the cavity floor at the same height as P1 and P2. Since the inclusion of sub-cavities alters the cavity geometry, the coordinates of these internal probes (labeled i–vii) vary across configurations. Table~\ref{tab:table2} lists the exact coordinates of these probes for each case.  
\begin{table}[ht]
    \centering
    \caption{\label{tab:table3}Comparisons of frequency modes obtained for various configurations for $M_\infty=2$.}
    \vspace{0.5em}
    \small\textit{Modes obtained from the analytical solution:} 
    \textit{0.23, 0.54, 0.89, 1.16, 1.46}.
    \vspace{0.5em}
    \begin{ruledtabular}
    \begin{tabular}{ccc}
\hline
        \multicolumn{3}{c}{\textbf{$M_\infty$=2}} \\
        \hline
        \textbf{Case} & \textbf{Modes (St)} & \textbf{Dominant Frequency (kHz)} \\
        \hline
        R2 & \makecell {0.283, \textbf{0.55}, 0.834,\\ 1.189, 1.36 }& 7.67\\
       F2 & \makecell{\textbf{0.68}, 0.97,1.14,\\ 1.28,1.36} & 9.62 \\
A2 & \makecell{ \textbf{0.52}, 0.62,0.80,\\ 1.14} & 7.36 \\

    \end{tabular}
    \end{ruledtabular}
\end{table}
 \begin{table}[ht]
    \centering
    \caption{\label{tab:table4}Comparisons of frequency modes obtained for various configurations for $M_\infty=3$.}
    \vspace{0.5em}
    \small\textit{Modes obtained from the analytical solution:} 
    \textit{0.21, 0.49, 0.78}.
    \vspace{0.5em}
    \begin{ruledtabular}
    \begin{tabular}{ccc}
     \hline
        \multicolumn{3}{c}{\textbf{$M_\infty$=3}} \\
        \hline
        \textbf{Case} & \textbf{Mode(St)} & \textbf{Dominant Frequency (kHz)} \\
        \hline
        R3 & \makecell{ 0.12, 0.21,0.44, 0.67, \textbf{0.82}} & 17.424 \\
    F3 & \makecell{\textbf{0.63}} & 13.39 \\
A3 & \makecell{0.21, 0.38,0.48,0.57,\textbf{0.75} } & 15.94 \\
    \end{tabular}
    \end{ruledtabular}
\end{table}

Fig. \ref{fig:2} illustrates the midsection of the geometry (x-y plane) of the confined cavity. The origin (0,0) lies at the leading edge of the cavity. As shown in Fig. \ref{fig:2}, we designate the shock's origin at the top wall's deflection corner as "a" and its impingement point on the shear layer as "b" to assess the shock's origin and its variation with flow conditions. The red line represents the impinging shock. Point 'b' provides a preliminary estimate of the approximate impingement distance from the leading edge. In our study, we have considered two $M_\infty$, namely 2 and 3. According to the $\theta$-$\beta$-$M_\infty$ relation, increasing $M_\infty$ with a fixed $\theta$ decreases $\beta$. In this analysis, we set the deflection angle to $2.29^\circ$ and observe the reduction in $\beta$ as $M_\infty$ increases. As $M_\infty$ increases, point 'b' moves downstream. Therefore, the impinging point of the shock on the shear layer is more downstream in the case of $M_\infty$ =3 than $M_\infty$ =2. We have already discussed the impact of this downstream movement of the impinging point in our recent study \cite{bhaduri2025influence}, and this paper will refer to it accordingly.
 \subsection{Governing equations}
We employ Large Eddy Simulations (LES) in this study to investigate compressible flow properties. To capture the essential physics, we solve the Favre-averaged filtered Navier–Stokes equations (Eqs.~\ref{eq:continuity}, \ref{eq:momentum}, and \ref{eq:energy})~\cite{moin1982numerical,piomelli1999large,cant2001sb,george2013lectures}.  The working fluid behaves as an ideal gas. The governing equations~\cite{soni2019modal,arya2021effect} used in the present LES framework are presented below:
\begin{equation}
\frac{\partial \overline{\rho}}{\partial t} +
\frac{\partial}{\partial x_i}\left[ \overline{\rho} \widetilde{u_i} \right] = 0
\label{eq:continuity}
\end{equation}

\begin{equation}
\frac{\partial}{\partial t} \left( \overline{\rho} \widetilde{u_i} \right) +
\frac{\partial}{\partial x_j} \left( \overline{\rho} \widetilde{u_i} \widetilde{u_j} \right) =
- \frac{\partial}{\partial x_i} \left( \overline{p} \right) +
\frac{\partial}{\partial x_j} \left[ (\mu + \mu_t) \frac{\partial \widetilde{u_i}}{\partial x_j} \right]   \label{eq:momentum}
\end{equation}

\begin{align}
  \frac{\partial}{\partial t} 
  \left( \overline{\rho} \widetilde{E} \right) +
  \frac{\partial}{\partial x_i} 
  \left( \overline{\rho} \widetilde{u_i} \widetilde{E} \right) &= 
  - \frac{\partial}{\partial x_j} 
  \left[ \widetilde{u_j} \left( - \widetilde{p} I + (\mu + \mu_t) 
  \frac{\partial \widetilde{u_i}}{\partial x_j} \right) \right] \notag \\ 
  &\quad + \frac{\partial}{\partial x_i} 
  \left[ \left( k + \mu_t C_p \text{Pr}_t \right) \frac{\partial \widetilde{T}}{\partial x_i} \right]
  \label{eq:energy}
\end{align}

where \( \overline{(.)} \) denotes time-averaged parameters, and \( \widetilde{(.)} \) represents density-averaged parameters.$\rho$ is the density, $u_i$ is the velocity vector, $p$ is the pressure and 
$E = e + \frac{u_i^2}{2}$ is the total energy of the system, where $e = \frac{h - p}{\rho}$ 
is the internal energy; and $h$ is the enthalpy.  
$\mu$ and $k$ represent the viscosity and thermal conductivity of the fluid, respectively, 
and $\mu_t$ and $\text{Pr}_t$ denote the  eddy viscosity and 
the turbulent Prandtl number, respectively.
  
The simulations resolve the large-scale turbulent eddies directly, while the Dynamic Wall-Adapting Local Eddy-viscosity (Dynamic WALE) model~\cite{nicoud1999subgrid, toda2010dynamic,arya2019effect} computes the eddy viscosity \(\mu_t\) to represent the effects of unresolved small-scale eddies (Eq.~\ref{eq:mu}). 
 \begin{equation} \label{eq:mu}
\mu_{t} = \rho \left( C_w \cdot \Delta_s \right)^2 
\frac{\left( S_{ij}^{d} S_{ij}^{d} \right)^{3/2}}
{\left( \widetilde{S}_{ij} \widetilde{S}_{ij} \right)^{5/2} + \left( S_{ij}^{d} S_{ij}^{d} \right)^{5/4}}
\end{equation}
where $\widetilde{S}_{ij} = \frac{1}{2} (\widetilde{g}_{ij} + \widetilde{g}_{ji} ), \quad    \widetilde{g}_{ij} = \frac{\partial \widetilde{u}_i}{\partial x_j}$ and ($\Delta_s$) is the filter width. \ 
$C_w$ is a model constant that is calculated dynamically based on the proximity to the wall.

Unlike the standard WALE model, which uses a fixed model constant, the dynamic WALE approach adapts this constant using the Germano–Lilly procedure. However, Toda et al.~\cite{toda2010dynamic} reported that this method tends to overestimate the eddy viscosity near walls due to elevated values of the WALE constant. Since the molecular viscosity dominates in the near-wall region, it is not necessary to dynamically compute the model constant if the model incorporates proper wall behavior. Consequently, they introduced a \textit{Shear and Vortex Sensor (SVS)} that detects the proximity to walls, defined as
\[
\text{SVS} = \frac{\left(S_{ij}^d S_{ij}^d\right)^{3/2}}{\left(\tilde{S}_{ij} \tilde{S}_{ij}\right)^3 + \left(S_{ij}^d S_{ij}^d\right)^{3/2}},
\]
which approaches zero in pure shear and one in pure rotation. The dynamic procedure to calculate $C_w$ is applied only where \(\text{SVS} > 0.09\), and a fixed WALE constant of 0.5 is used where \(\text{SVS} \leq 0.09\), thereby preserving accurate near-wall behavior.

\subsection{Numerical schemes and Validation.}
To solve the governing equations, we use rhoCentralRK4Foam ~\cite{li2020scalability} in the OpenFOAM ~\cite{jasak2007openfoam}  framework. For explicit time integration with third-order accuracy, this density-based solver, which is based on the finite volume approach, uses a four-stage, low-storage Runge–Kutta scheme. Kurganov and Tadmor's central upwind scheme \cite{kurganov2000new,adityanarayan2023leading}and the central scheme for the dissipative fluxes are used to discretize the convective fluxes. The specific heat of air is provided by the Joint Army-Navy-Air Force (JANAF) thermochemical tables \cite{stull1974janaf} , whereas Sutherland's law predicts the temperature-dependent viscosity.

We have validated the solver using the experimental data provided by Gruber et al.~\cite{gruber2001fundamental}. The experiments were performed on an open cavity with a length-to-depth ratio of $L/D = 3$ at a freestream Mach number of $M_\infty = 3$, corresponding to a stagnation pressure of 690~kPa and a stagnation temperature of 300~K. Figure~\ref{fig:3} shows the comparison between the present numerical simulations and the experimental results. The time-averaged pressure ($\langle p \rangle$), normalized by the freestream pressure, along the cavity walls shows good agreement with the experimental data, within a tolerance of $\pm 5\%$.
  For a detailed discussion on solver validation, grid independence analysis, and LES verification, readers may refer to our recent work~\cite{bhaduri2024effects}, as we do not reiterate these topics here for brevity.The present study examines two freestream Mach numbers, \(M_\infty = 2\) and \(M_\infty = 3\), which correspond to velocities of 551.576~m/s and 828.683~m/s, respectively. We initialize the flow with ambient pressure and a static temperature of 189.29 K across all configurations. To introduce velocity fluctuations at the inlet, we employ the Klein inflow generator~\cite{klein2003digital}. We impose no-slip boundary conditions on all insulated walls, and at the outlet, we extrapolate all flow variables under the assumption of supersonic outflow.

 \begin{figure*}

	\includegraphics[scale=0.8]{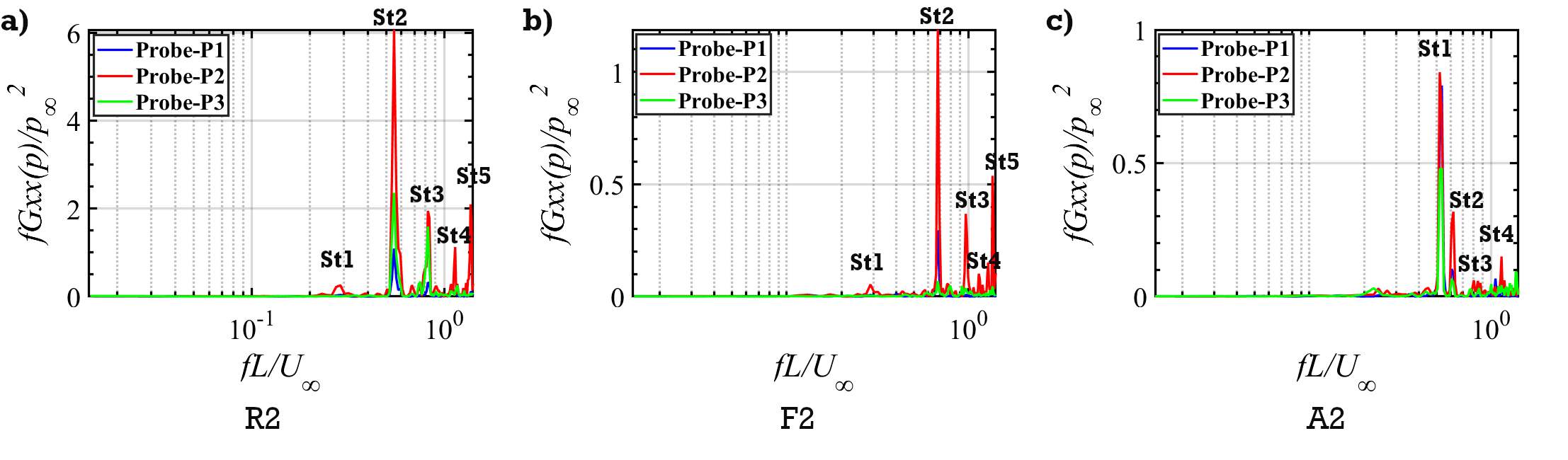}
    \centering

    \caption{\label{fig:4}Normalized Power Spectral Density (PSD) (fGxx(p)/$(p_\infty)^2$) vs the Strouhal number (St=fL/$U_\infty$)obtained from the pressure fluctuation data recorded by probes P1, P2 and P3 for cavity (a) without sub-cavity (R2) (b) with front-wall sub-cavity (F2) (c) with aft-wall sub-cavity (A2)  at M$_\infty$ = 2.  }
   \end{figure*}
   \begin{figure*}

	\includegraphics[scale=0.8]{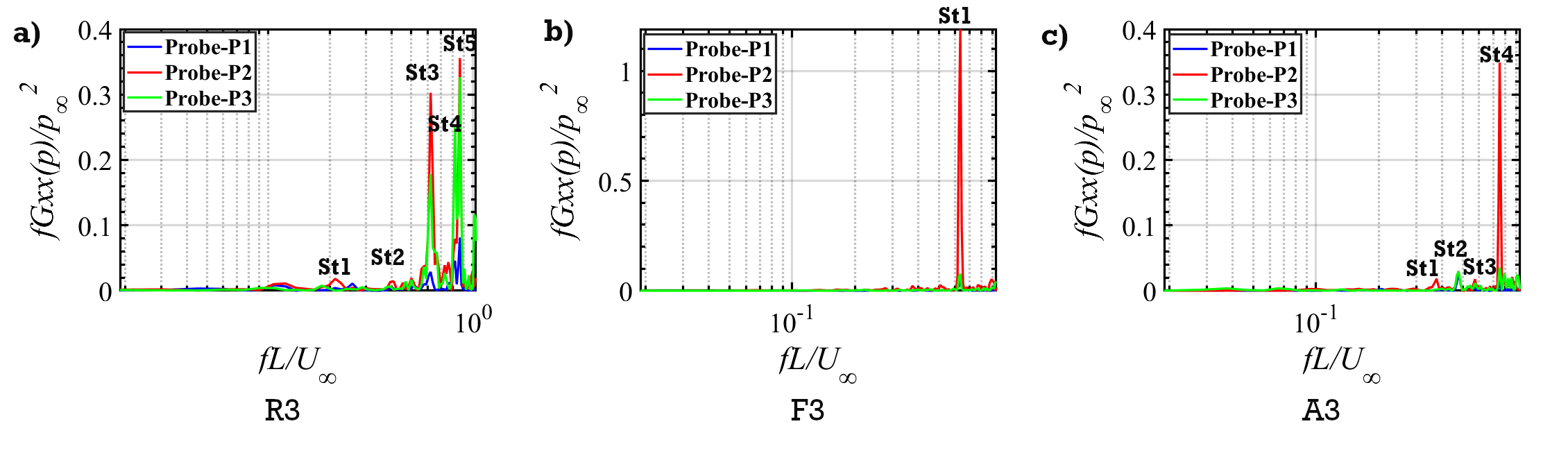}
    \centering

    \caption{\label{fig:5}Normalized Power Spectral Density (PSD) (fGxx(p)/$(p_\infty)^2$) vs the Strouhal number (St=fL/$U_\infty$) obtained from the pressure fluctuation data recorded by probes P1, P2 and P3 for cavity (a) without sub cavity (R3) (b) with front-wall sub-cavity (F3) (c) with aft-wall sub-cavity (A3) at M$_\infty$ = 3 }
   \end{figure*}
 \begin{figure*}[htbp]
    \centering
    \includegraphics[scale=1.2]{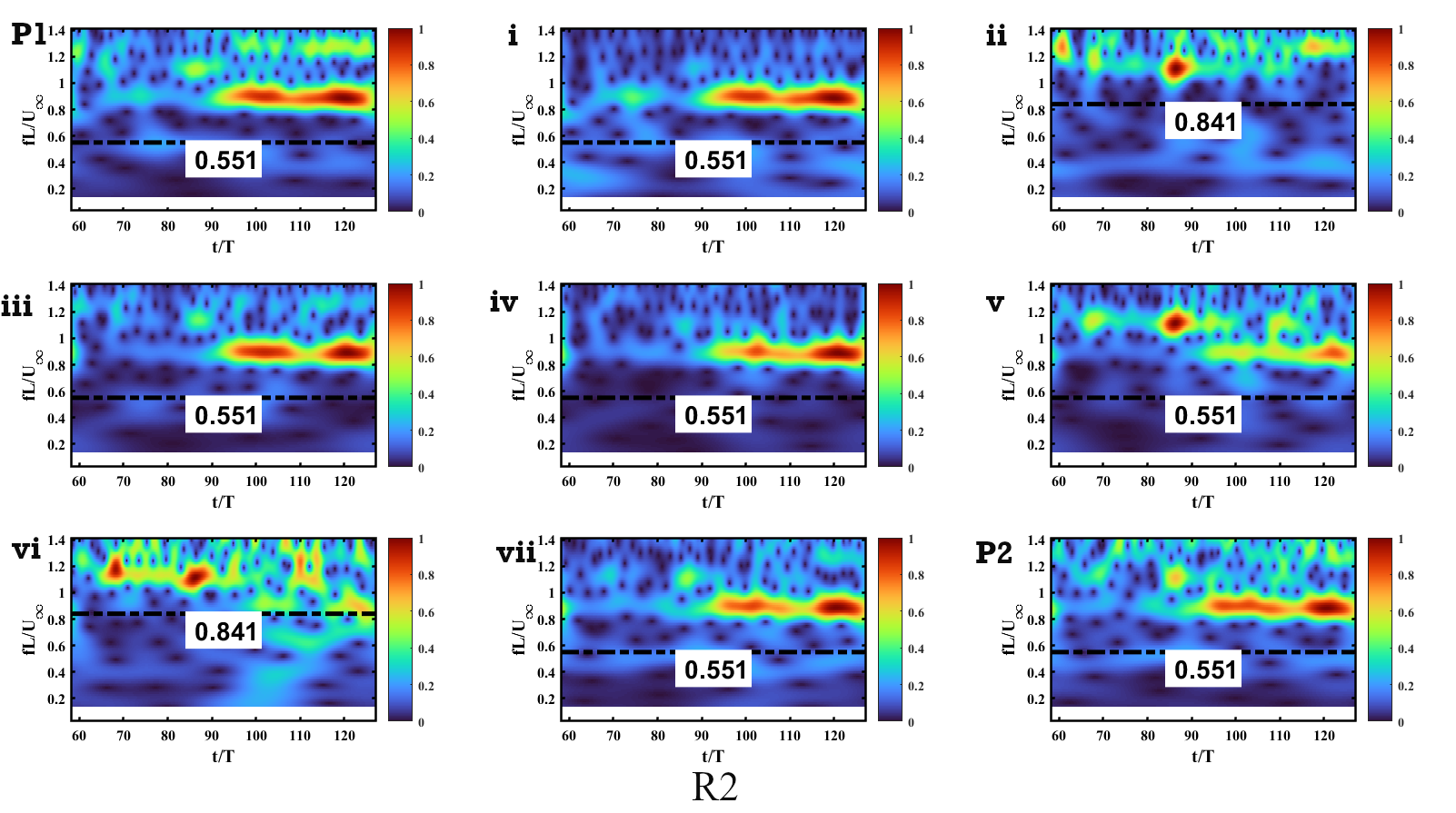}
    \caption{\label{fig:6} Normalized Continuous Wavelet Transform of the pressure signals obtained from P1, P2 and the internal probes i-vii for the reference case (R2) at $M_\infty = 2$ .}
\end{figure*}

\begin{figure*}[htbp]
    \centering
    \includegraphics[scale=1.2]{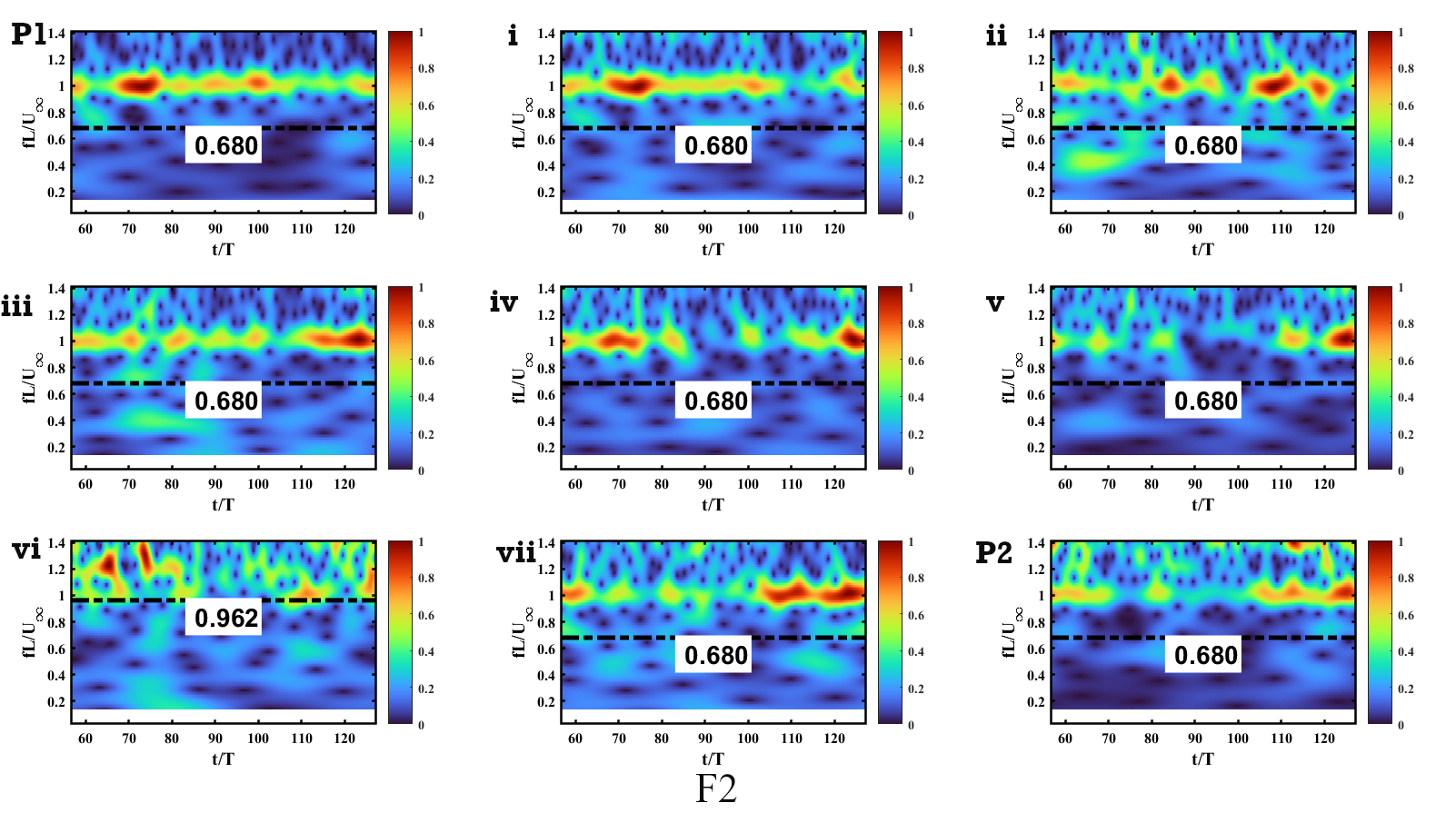}
    \caption{\label{fig:7} Normalized Continuous Wavelet Transform of the pressure signals obtained from P1, P2 and the internal probes i-vii for the front-wall subcavity case (F2) at $M_\infty = 2$. }
\end{figure*}

\begin{figure*}[htbp]
    \centering
    \includegraphics[scale=1.2]{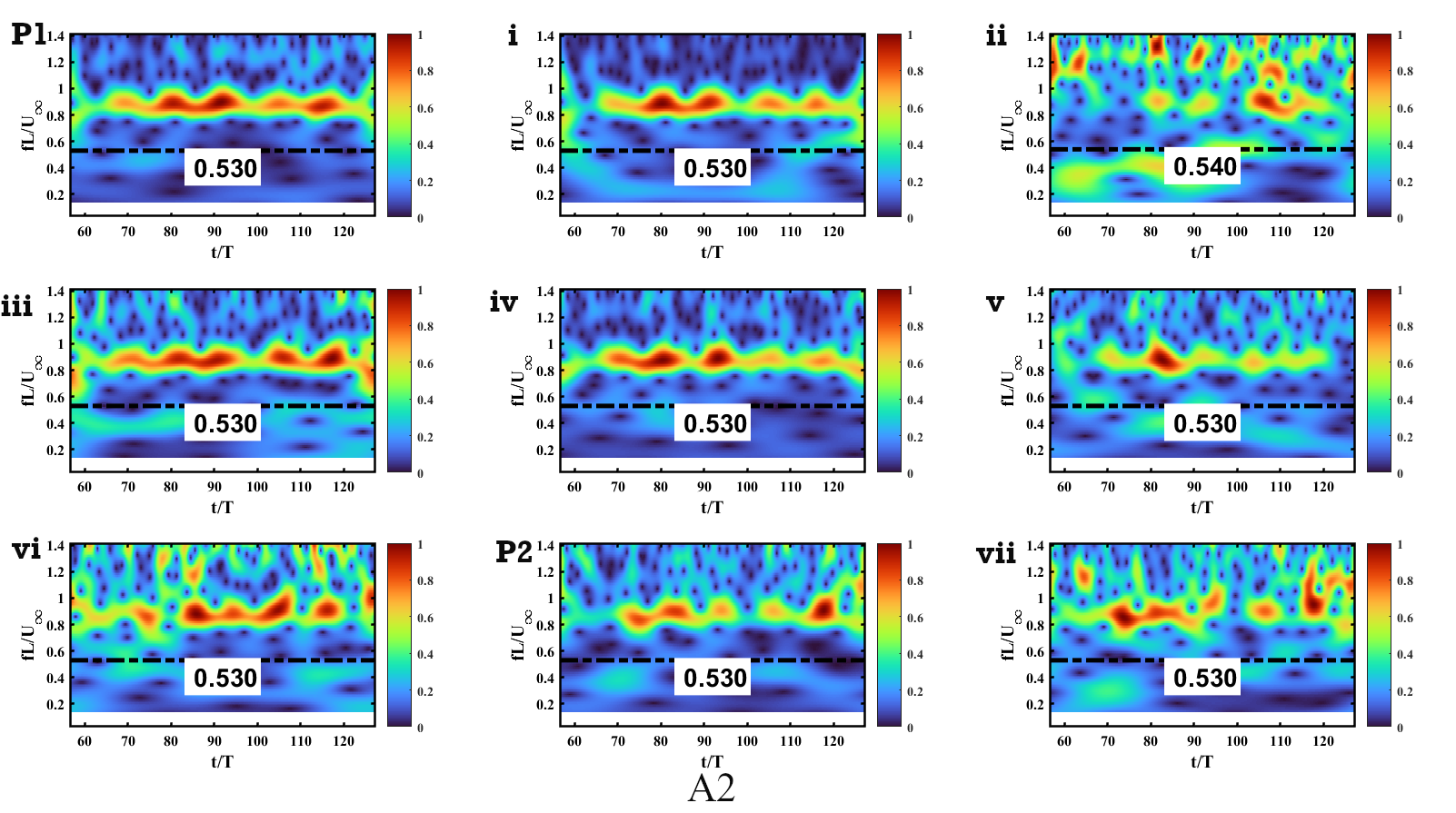}
    \caption{\label{fig:8} Normalized Continuous Wavelet Transform of the pressure signals obtained from P1, P2 and the internal probes i-vii for the aft-wall subcavity case (A2) at $M_\infty = 2$ .}
\end{figure*}
\begin{figure*}[htbp]
    \centering
    \includegraphics[scale=1]{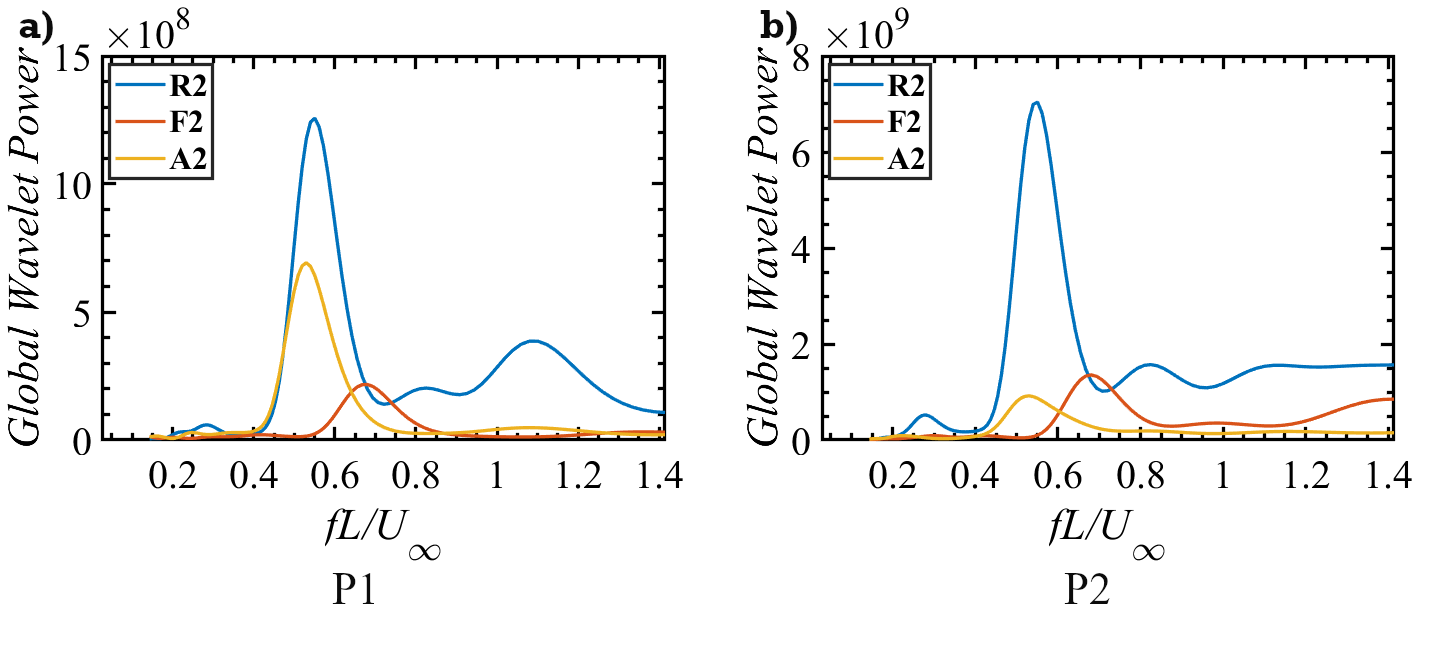}
    \caption{\label{fig:9} Global Wavelet Power  of the pressure signals obtained from P1 and P2 for $M_\infty = 2$ cases . }
\end{figure*}
\begin{figure*}[htbp]
    \centering
    \includegraphics[scale=1]{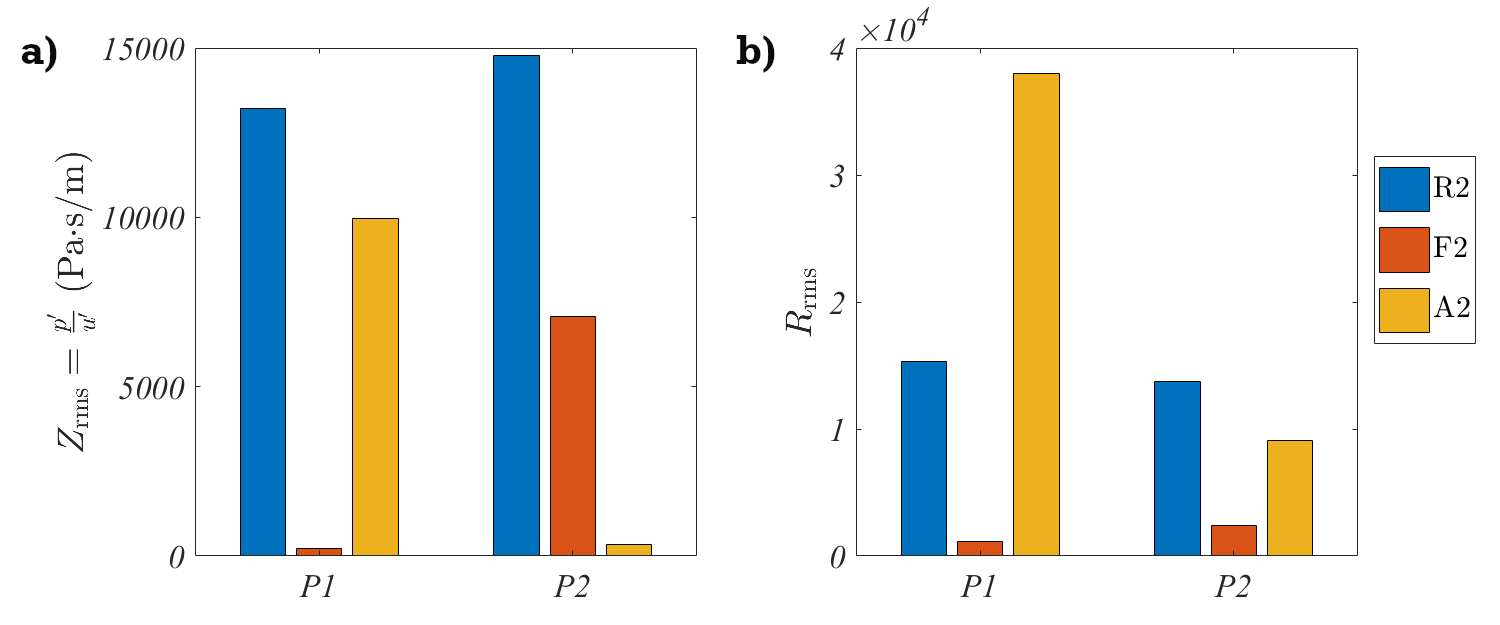}
    \caption{\label{fig:10} Acoustic impedance and normalized acoustic impedance of the pressure signals obtained from P1 and P2 for $M_\infty = 2$ cases.}
\end{figure*}
\begin{figure*}[htbp]
    \centering
    \includegraphics[scale=1.2]{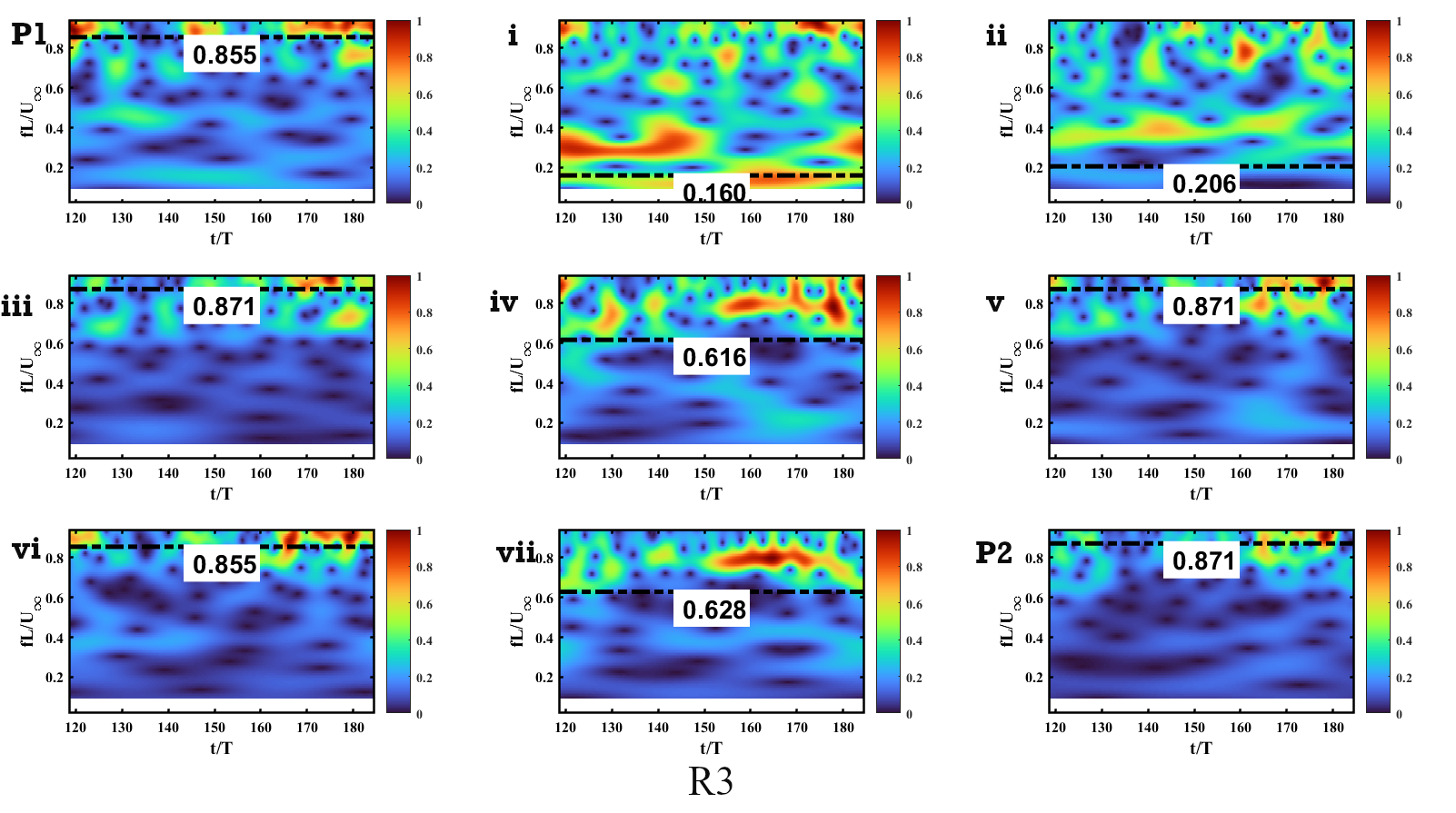}
    \caption{\label{fig:11} Normalized Continuous Wavelet Transform of the pressure signals obtained from P1, P2 and the internal probes i-vii for the reference case (R3) at $M_\infty = 3$ .}
\end{figure*}

\begin{figure*}[htbp]
    \centering
    \includegraphics[scale=1.2]{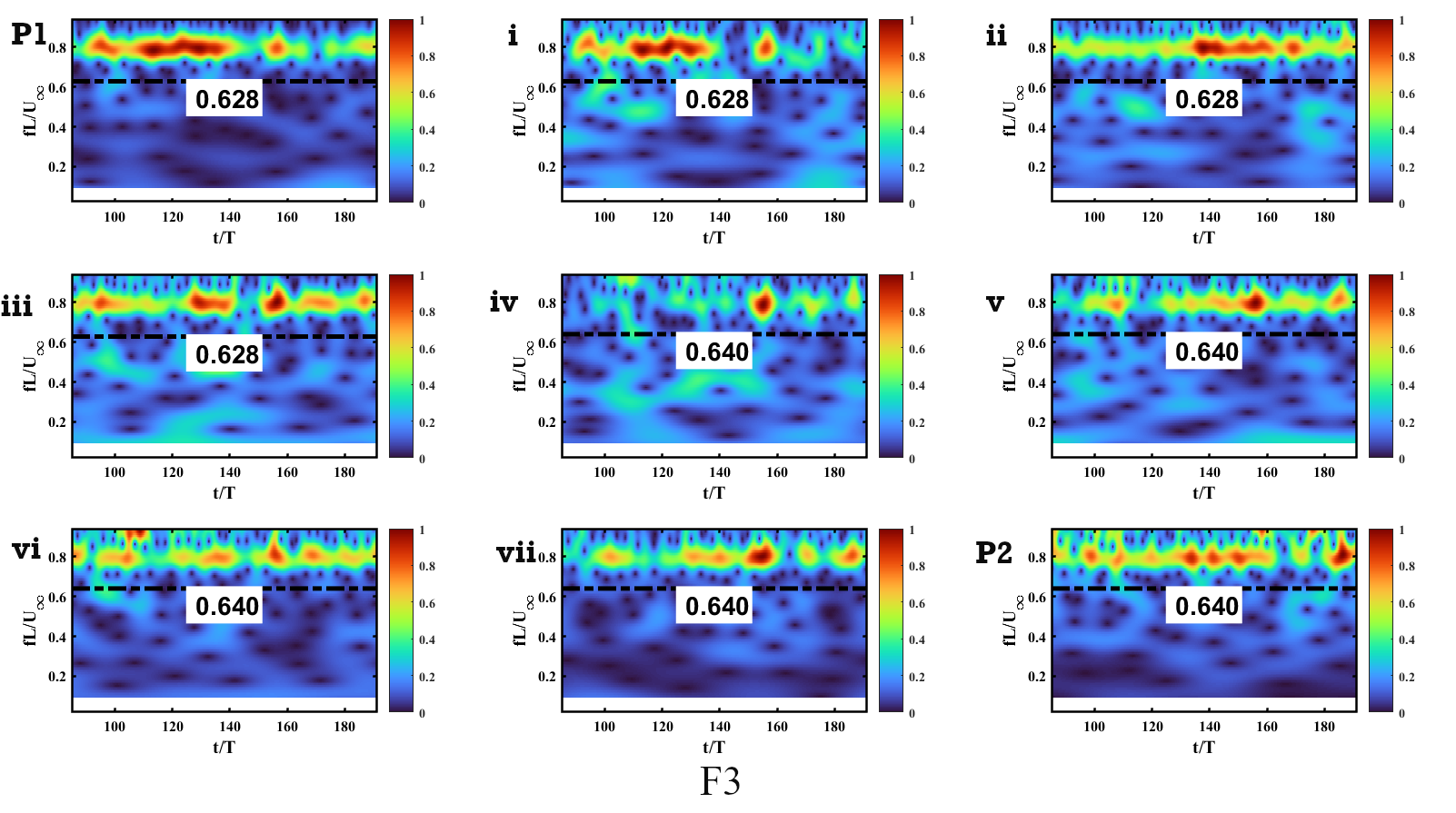}
    \caption{\label{fig:12} Normalized Continuous Wavelet Transform of the pressure signals obtained from P1, P2 and the internal probes i-vii for the front-wall subcavity case (F3) at $M_\infty = 3$.}
\end{figure*}

\begin{figure*}[htbp]
    \centering
    \includegraphics[scale=1.2]{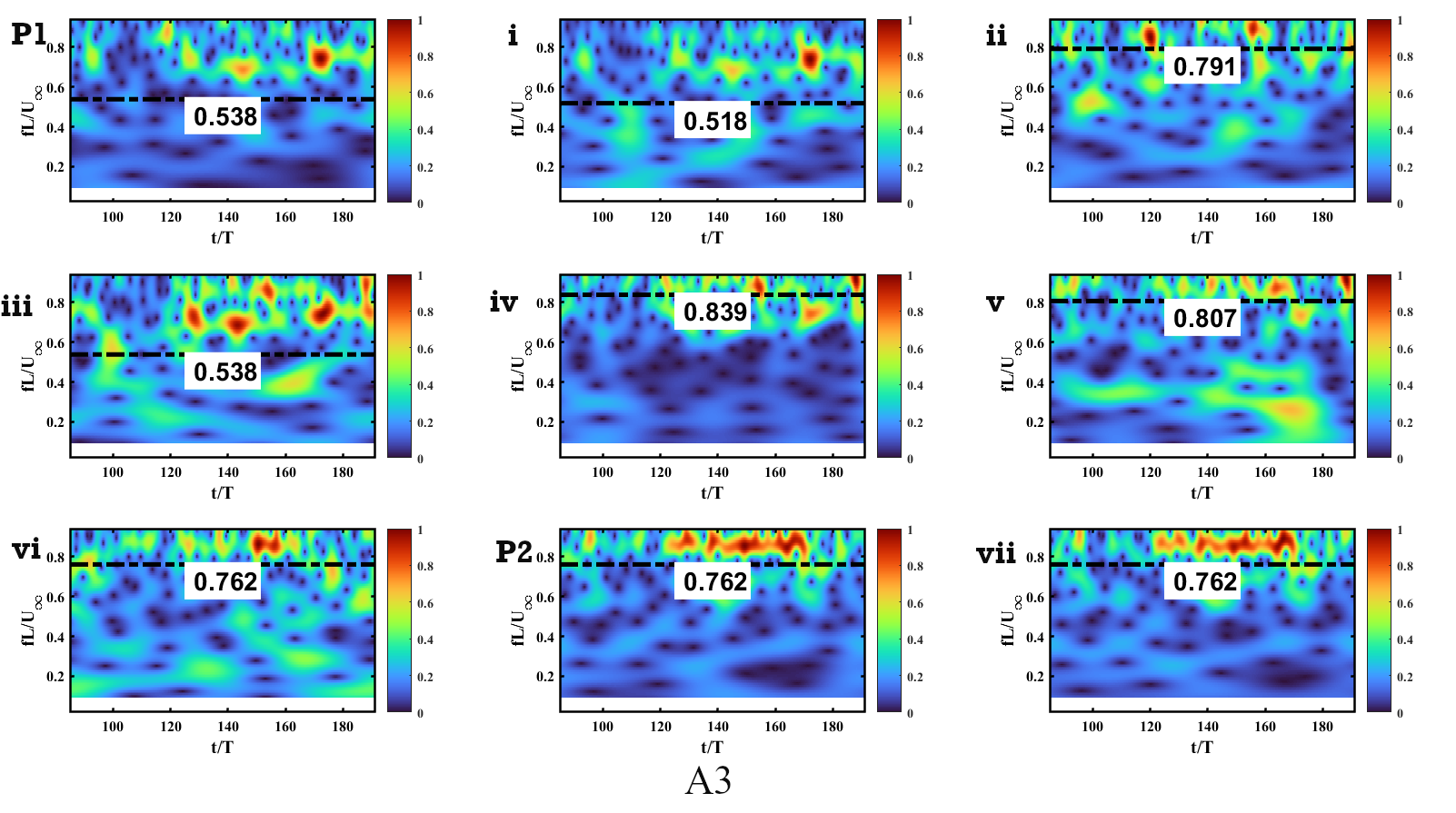}
    \caption{\label{fig:13} Normalized Continuous Wavelet Transform of the pressure signals obtained from P1, P2 and the internal probes i-vii for the aft-wall subcavity case (A3) at $M_\infty = 3$.}
\end{figure*}

\begin{figure*}[htbp]
    \centering
    \includegraphics[scale=1]{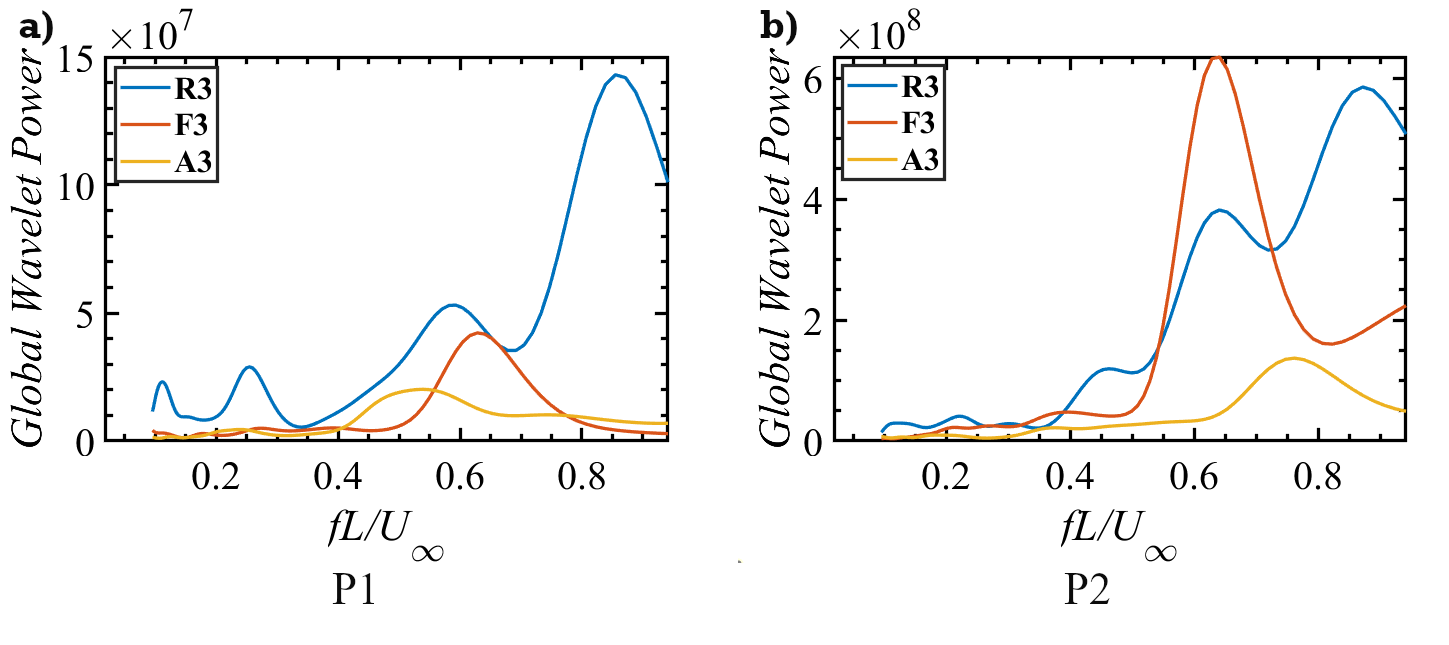}
    \caption{\label{fig:14} Global Wavelet Power  of the pressure signals obtained from P1 and P2 for $M_\infty = 3$ cases .}
\end{figure*}

\begin{figure*}[htbp]
    \centering
    \includegraphics[scale=1]{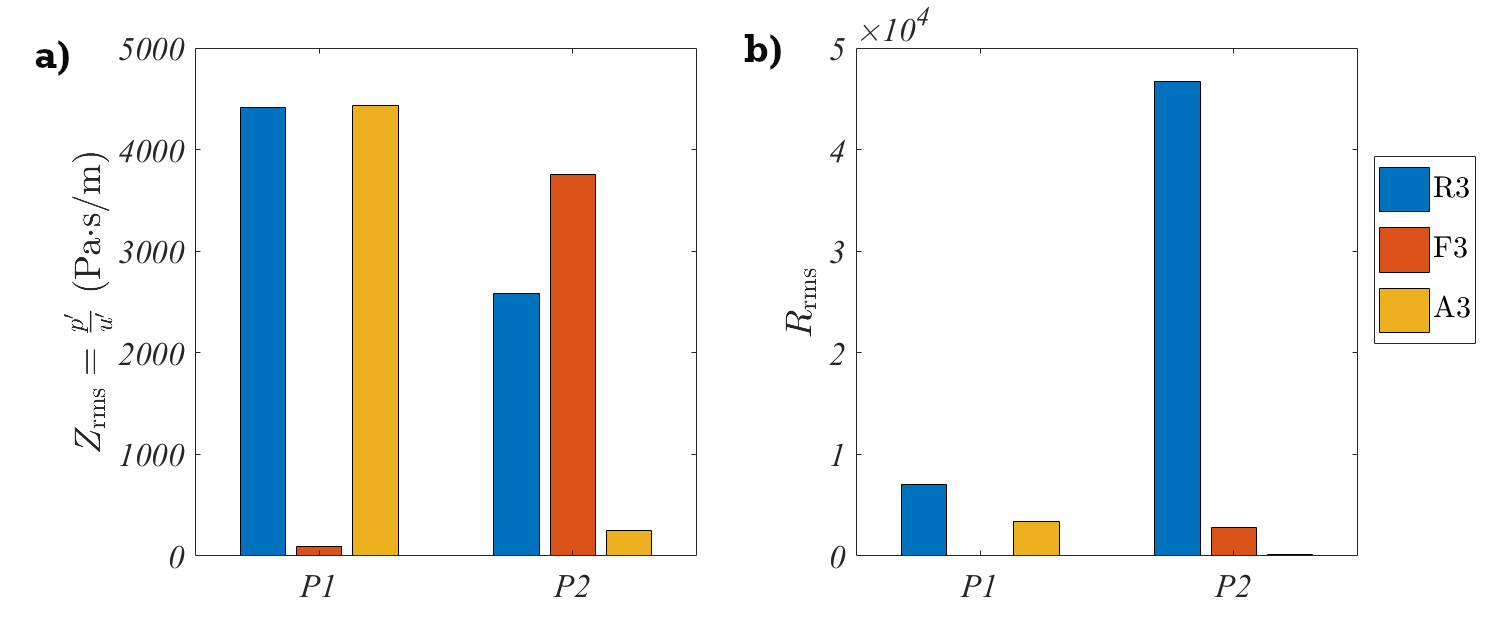}
    \caption{\label{fig:15} Acoustic impedance and normalized acoustic impedance of the pressure signals obtained from P1 and P2 for $M_\infty = 3$ cases.}
\end{figure*}
  
   \begin{figure*}

	\includegraphics[scale=1.05]{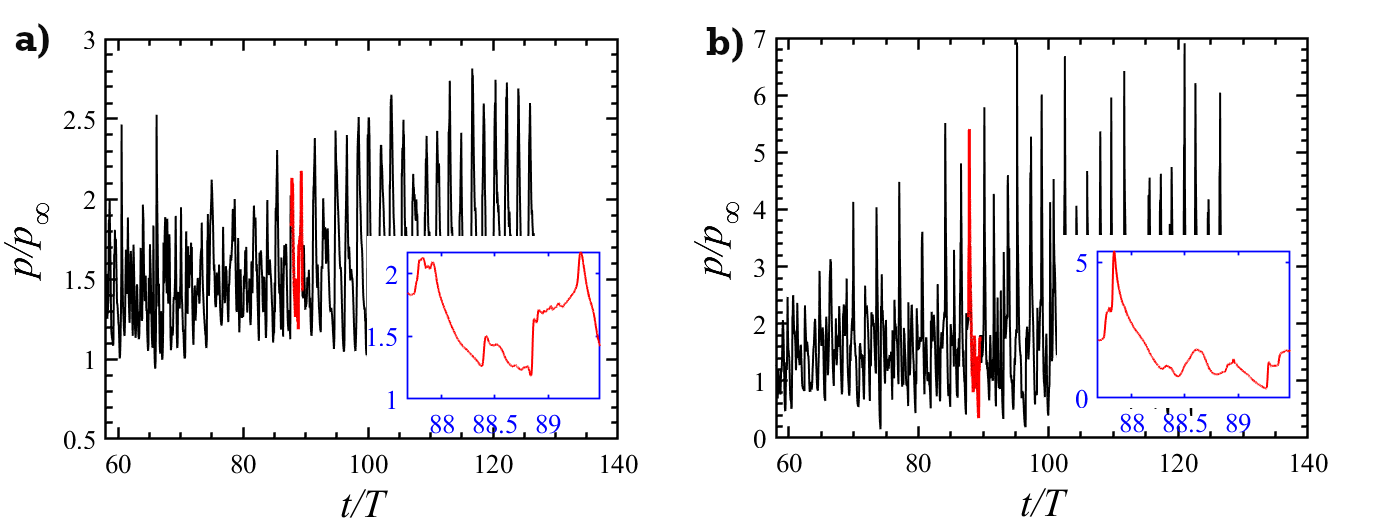}
    \centering

    \caption{\label{fig:16} Temporal variation of pressure normalized with the freestream pressure (p/p$_\infty$) for cavity configuration without sub-cavity  at M$_\infty$ = 2 (R2) at the a) front and b) aft edges. }
   \end{figure*}
     \begin{figure*}

	\includegraphics[width=1\textwidth]{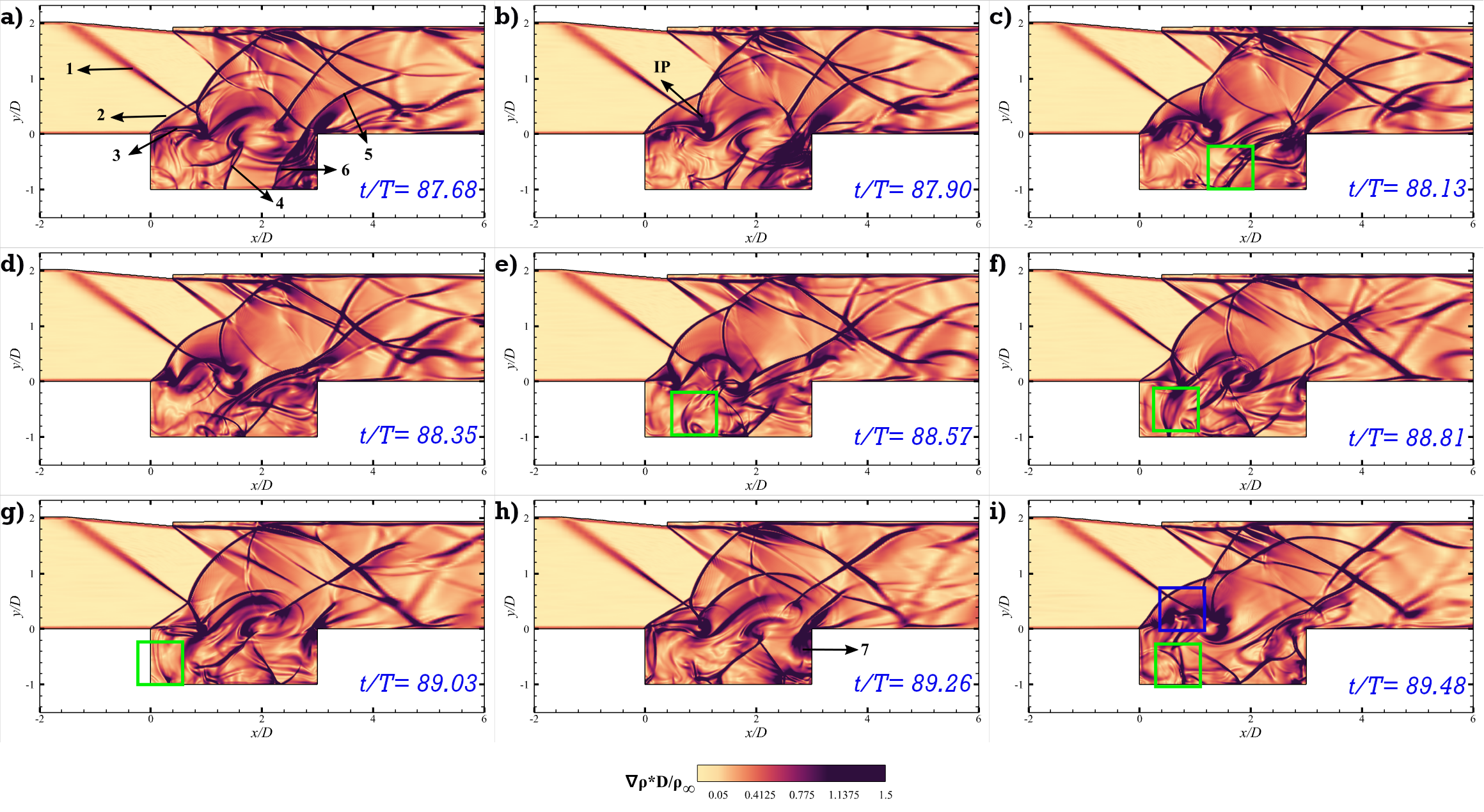}
    \centering

    \caption{\label{fig:17} Normalized Density Gradient contour ($\nabla \rho \cdot (D/\rho_{\infty})$) for one complete cycle from the time step (t/T) of 87.68 (a) to 89.48 (i) at an interval of 0.22 of the cavity conifguration without sub-cavity  at M$_\infty$ = 2 (R2). Key flow features: (1) the shock from the top wall deflection corner, (2) the leading edge separation shock, (3) the separating shear layer (4) the downstream traveling pressure wave, (5) reattachment shock (6) the upstream traveling pressure wave of the present cycle, and (7) pressure wave of the next cycle. IP is the impinging point. The green square marks the upstream traveling pressure wave. The blue square demonstrates the perturbations in the shear layer.}
   \end{figure*}
  
   \begin{figure*}

	\includegraphics[scale=1.05]{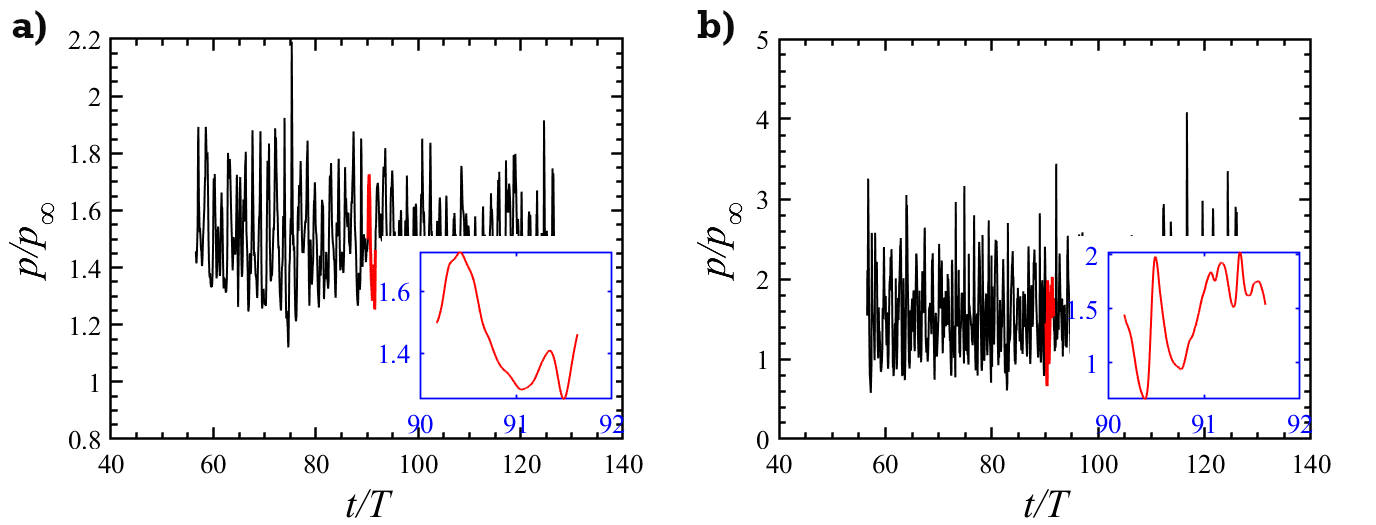}
    \centering

    \caption{\label{fig:18}Temporal variation of pressure normalized with the freestream pressure (p/p$_\infty$) for cavity configuration with front-wall sub-cavity  at M$_\infty$ = 2 (F2) at the a) front and b) aft edges. }
   \end{figure*}
   \begin{figure*}

	\includegraphics[width=1\textwidth]{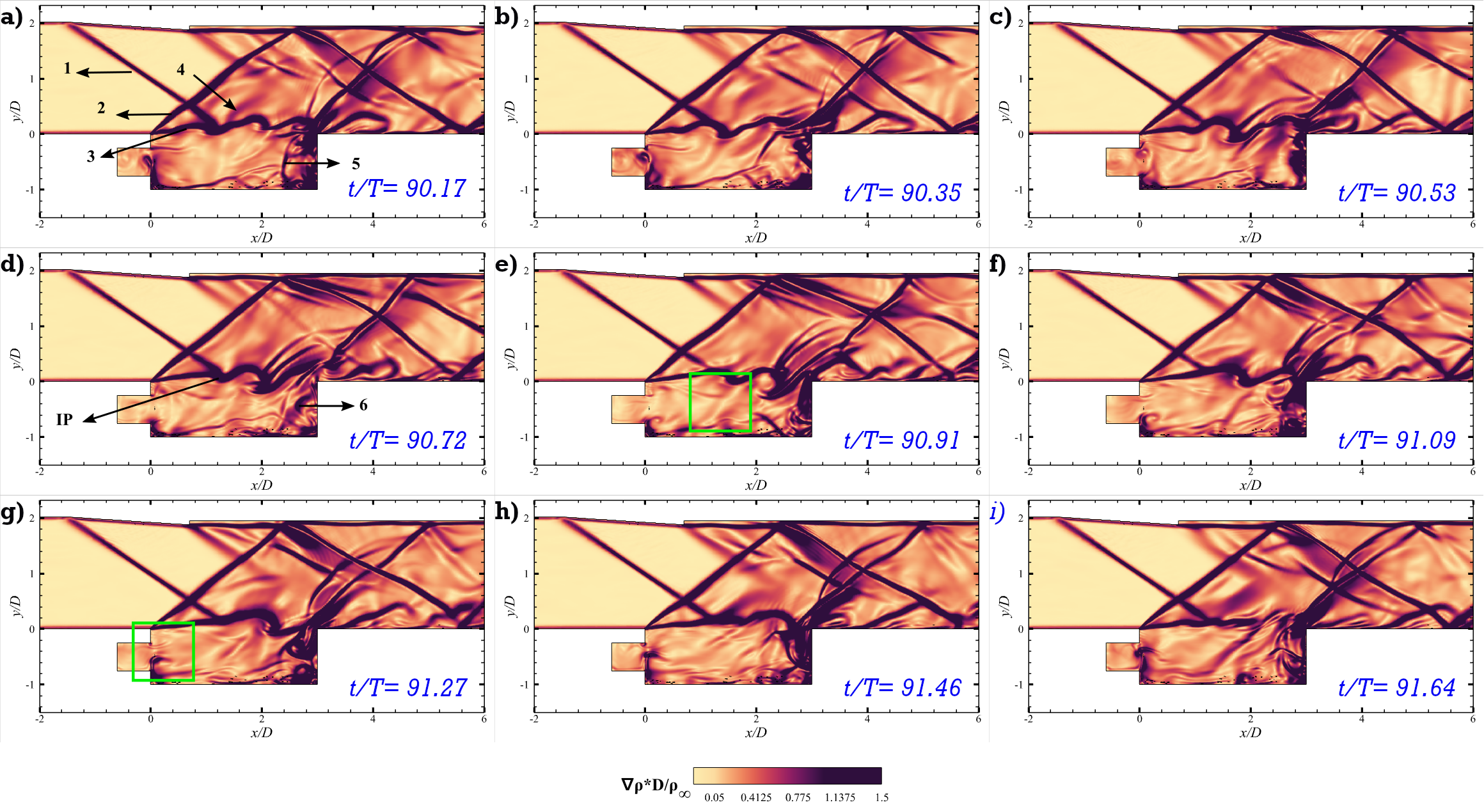}
    \centering

    \caption{\label{fig:19}Normalized Density Gradient contour ($\nabla \rho \cdot (D/\rho_{\infty})$) for one complete cycle from the time step (t/T) of 90.17 (a) to 91.64 (i) at an interval of 0.18 of the cavity configuration with front-wall sub-cavity at M$_\infty$ = 2 (F2). Key flow features: (1) the shock from the top wall deflection corner, (2) the leading edge separation shock, (3) the separating shear layer (4) expansion wave formed as the shear layer reflects the shock wave impinging on it (5) the upstream traveling pressure wave of the present cycle and (6) the upstream traveling pressure wave of the next cycle. IP is the impinging point. The green square marks the upstream traveling pressure wave.}
   \end{figure*}
   \begin{figure*}

	\includegraphics[scale=1.05]{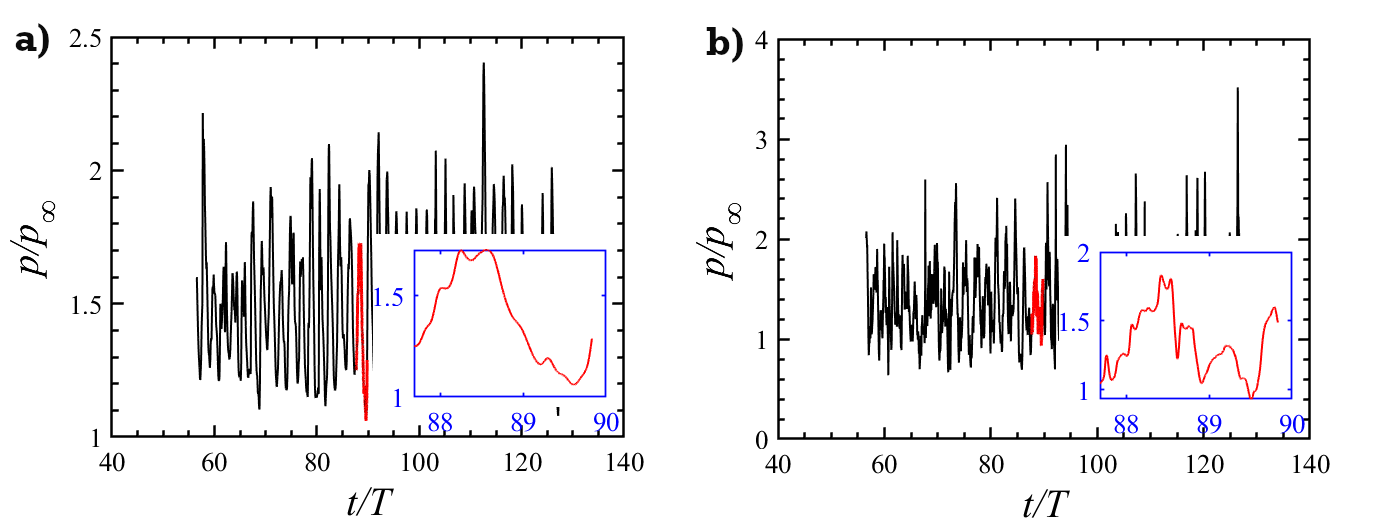}
    \centering

    \caption{\label{fig:20}Temporal variation of pressure normalized with the freestream pressure (p/p$_\infty$) for cavity configuration with aft-wall sub-cavity  at M$_\infty$ = 2 (A2) at the a) front and b) aft edges. }
   \end{figure*}
   \begin{figure*}

	\includegraphics[width=1\textwidth]{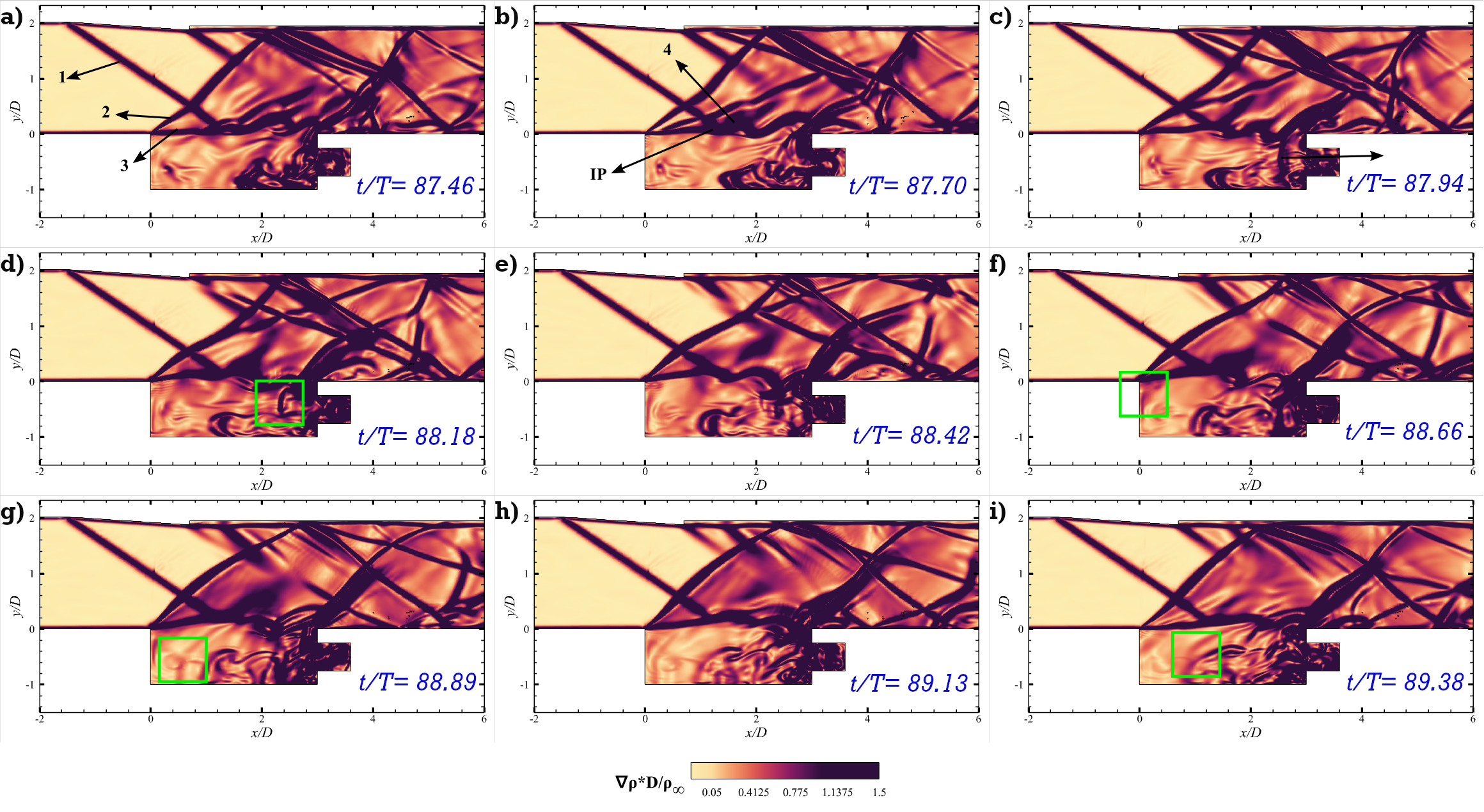}
    \centering

    \caption{\label{fig:21}Normalized Density Gradient contour ($\nabla \rho \cdot (D/\rho_{\infty})$) for one complete cycle from the time step (t/T) of 87.46 (a) to 89.38 (i) at an interval of 0.24 of the cavity configuration with aft-wall sub-cavity at M$_\infty$ = 2 (A2). Key flow features: (1) the shock from the top wall deflection corner, (2) the leading edge separation shock, (3) the separating shear layer (4) expansion wave formed as the shear layer reflects the shock wave impinging on it (5) the upstream traveling pressure wave of the present cycle. IP is the impinging point. The green square marks the upstream traveling pressure wave.}
   \end{figure*}

  \begin{figure*}

	\includegraphics[scale=1.05]{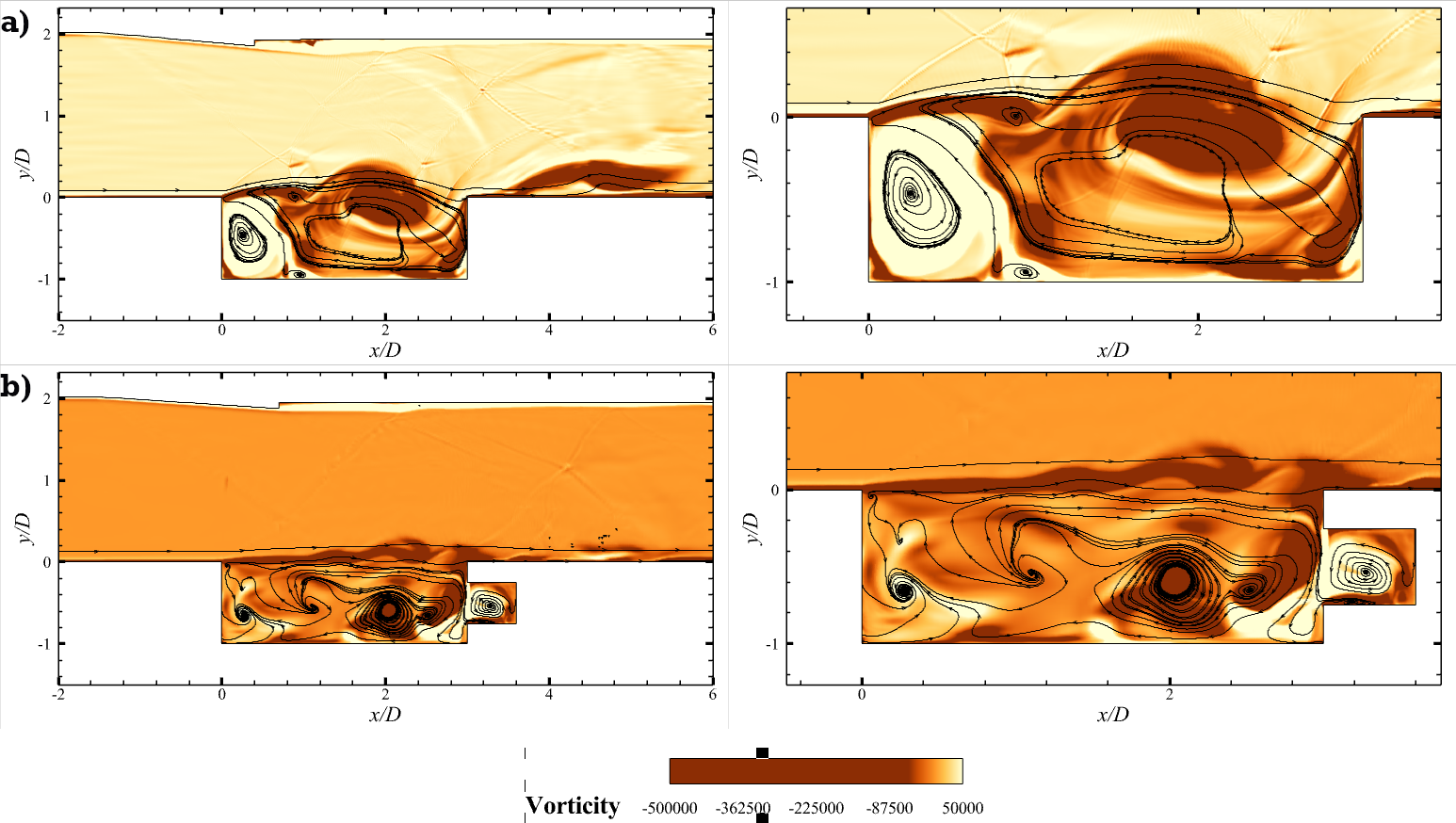}
    \centering

    \caption{\label{fig:22} Vorticity fields of the cavity a) without sub-cavity (R2) and b) with aft-wall sub-cavity (A2) at $M_\infty$=2. The close-up of the vorticity field with the streamlines near the aft edge of the cavity shows the difference in the flow field in the absence and presence of the sub-cavity at t/T= 87.68 and 87.46, respectively.}
   \end{figure*}
   \begin{figure*}

	\includegraphics[scale=1.1]{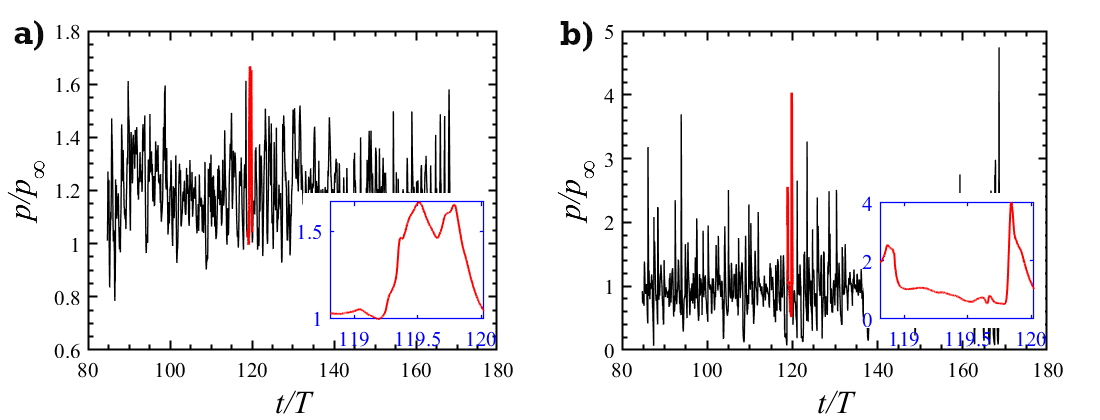}
    \centering

    \caption{\label{fig:23}Temporal variation of pressure normalized with the freestream pressure (p/p$_\infty$) for cavity configuration without sub-cavity at M$_\infty$ = 3 (R3) at the a) front and b) aft edges.}
   \end{figure*}
   \begin{figure*}

	\includegraphics[width=1\textwidth]{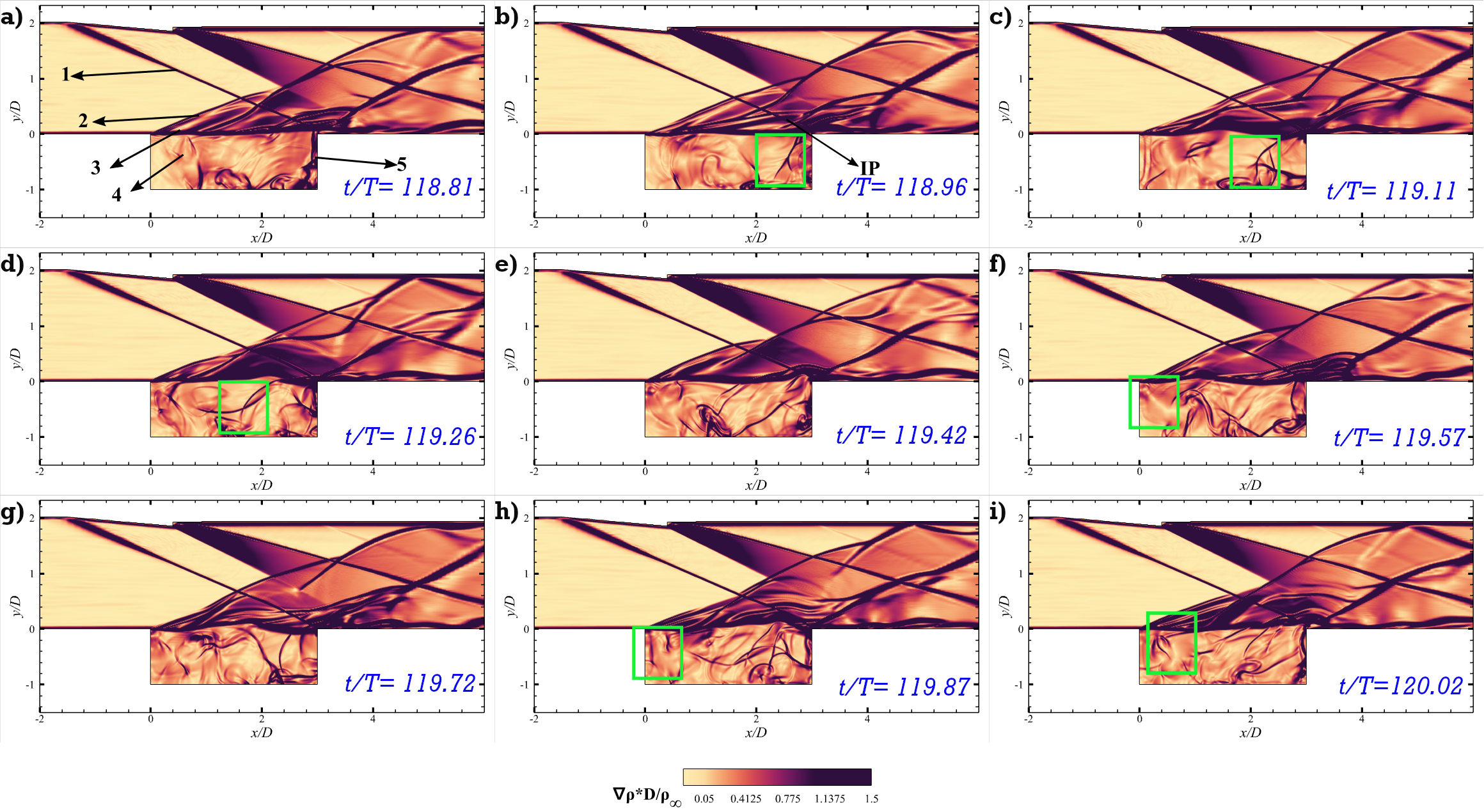}
    \centering

    \caption{\label{fig:24} Normalized Density Gradient contour ($\nabla \rho \cdot (D/\rho_{\infty})$) for one complete cycle from the time step (t/T) of 118.81 (a) to 120.02 (i) at an interval of 0.15 of the cavity configuration without sub-cavity at M$_\infty$ = 3 (R3). Key flow features: (1) the shock from the top wall deflection corner, (2) the leading edge separation shock, (3) the separating shear layer, (4) downstream traveling wave, (5) the upstream traveling pressure wave of the present cycle, and (6) the upstream traveling wave of the next cycle. IP is the impinging point. The green square marks the upstream traveling pressure wave. }
   \end{figure*}
   \begin{figure*}

	\includegraphics[scale=1.05]{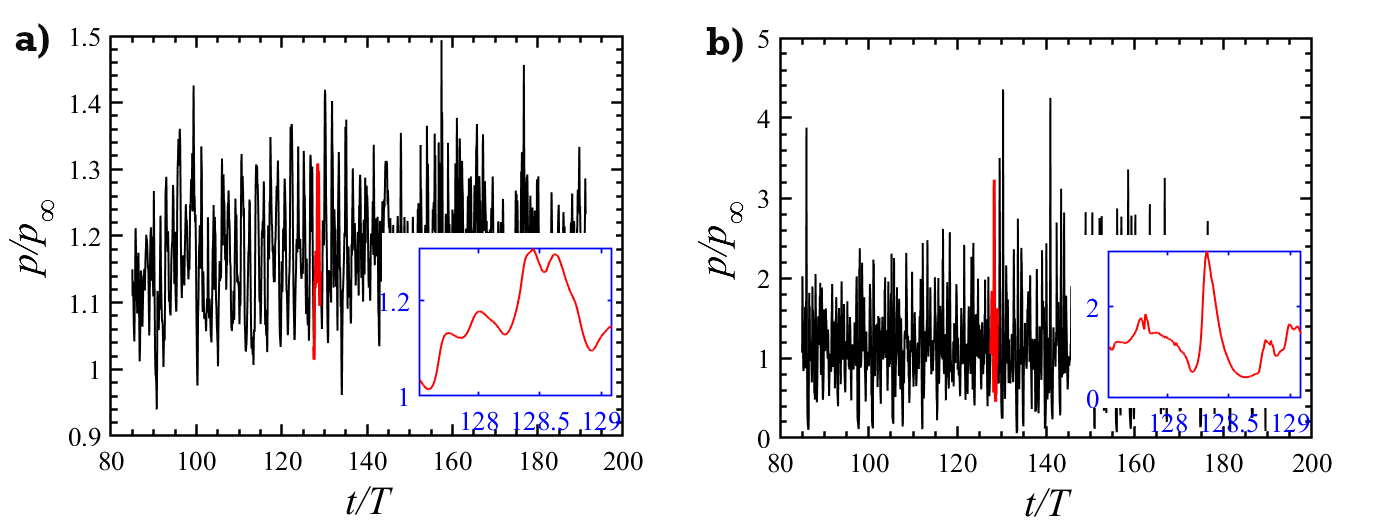}
    \centering

    \caption{\label{fig:25}Temporal variation of pressure normalized with the freestream pressure (p/p$_\infty$) for cavity configuration with front-wall sub-cavity at M$_\infty$ = 3 (F3) at the a) front and b) aft edges. }
   \end{figure*}
   \begin{figure*}

	\includegraphics[width=1\textwidth]{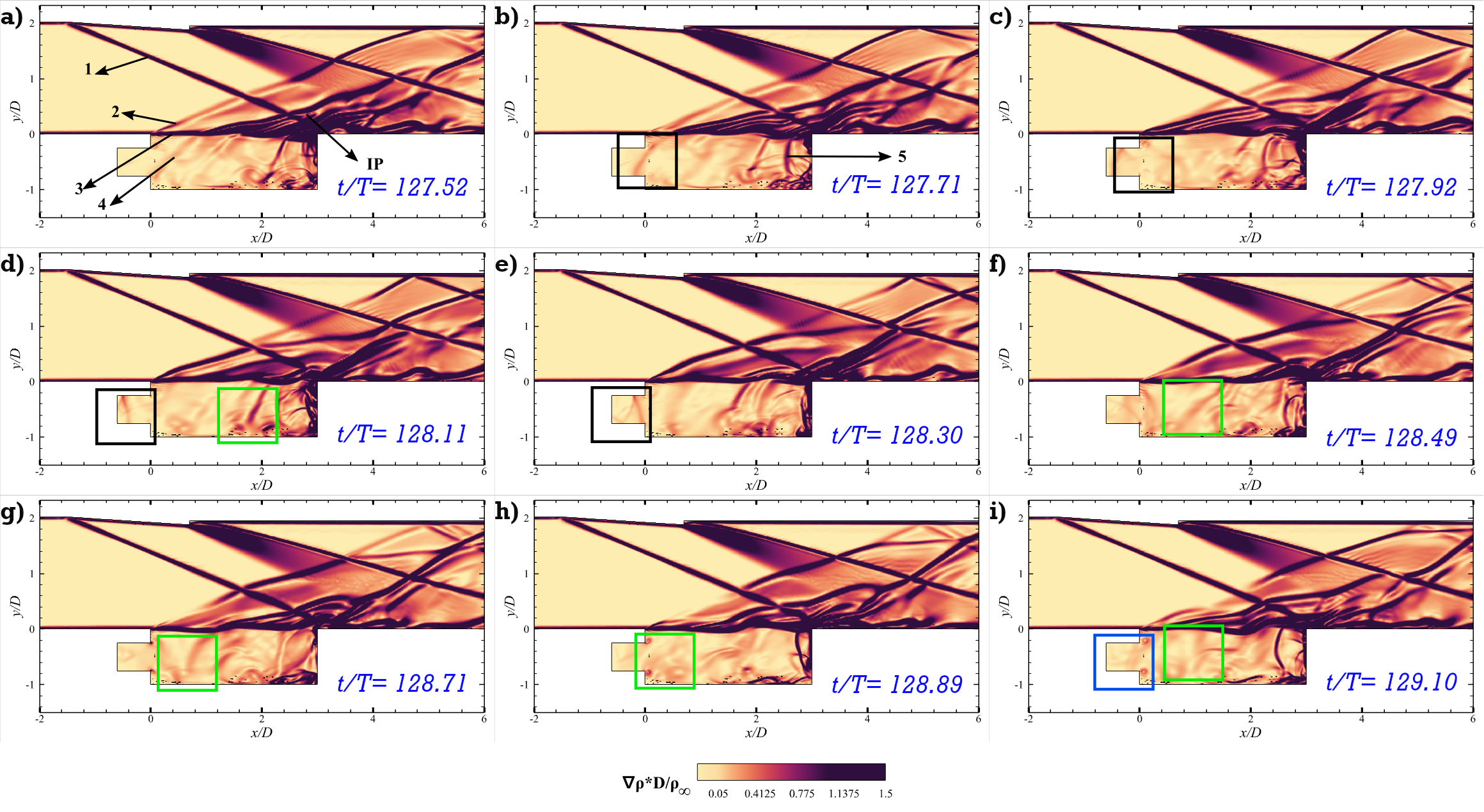}
    \centering

    \caption{\label{fig:26} Normalized Density Gradient contour ($\nabla \rho \cdot (D/\rho_{\infty})$) for one complete cycle from the time step (t/T) of 127.52 (a) to 129.10 (i) at an interval of 0.19 of the cavity configuration with front-wall sub-cavity at M$_\infty$ = 3 (F3). Key flow features: (1) the shock from the top wall deflection corner, (2) the leading edge separation shock, (3) the separating shear layer (4) the pressure wave of the previous cycle,(5) the upstream traveling pressure wave of the present cycle, (6) the upstream traveling pressure wave of the next cycle. IP is the impinging point. The black square traces the pressure wave of the previous cycle, which enters the sub-cavity and reflects from the wall of the sub-cavity. The green and blue squares mark the upstream traveling pressure wave of the present cycle.}
   \end{figure*}
  
   \begin{figure*}

	\includegraphics[scale=1.05]{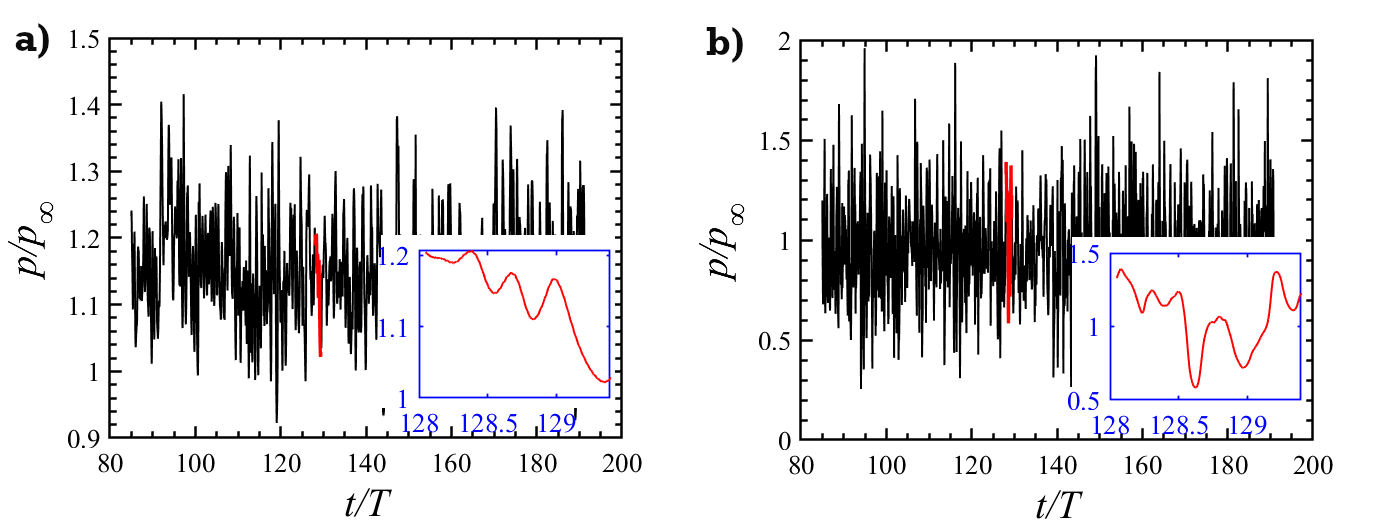}
    \centering

    \caption{\label{fig:27}Temporal variation of pressure normalized with the freestream pressure (p/p$_\infty$) for cavity configuration with aft-wall sub-cavity at M$_\infty$ = 3 (A3) at the a) front and b) aft edges . }
   \end{figure*}
   \begin{figure*}

    \includegraphics[width=1\textwidth]{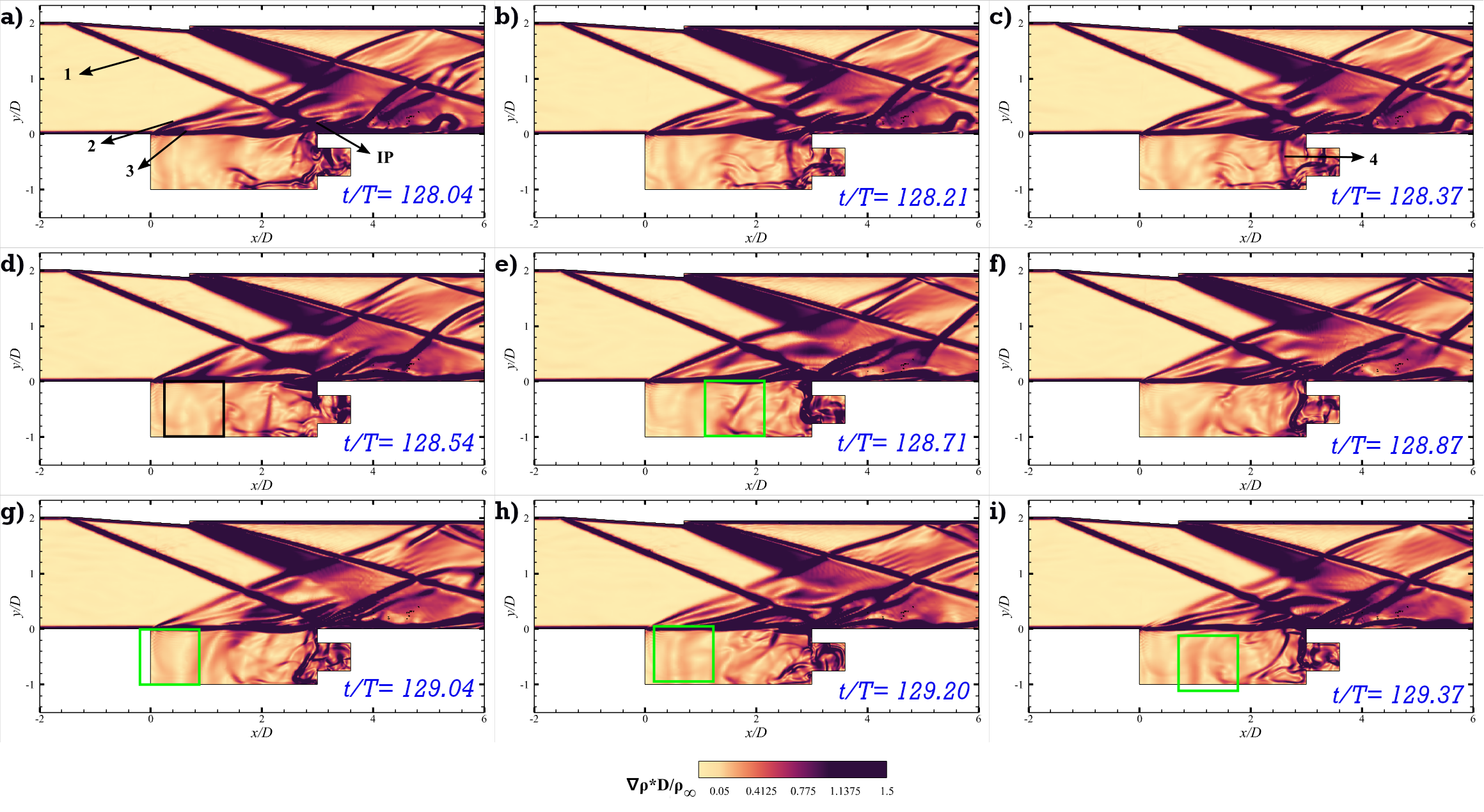}
    \centering

    \caption{\label{fig:28} Normalized Density Gradient contour ($\nabla \rho \cdot (D/\rho_{\infty})$) for one complete cycle from the time step (t/T) of 128.04 (a) to 129.37 (i) at an interval of 0.167 of the cavity configuration with aft-wall sub-cavity at M$_\infty$ = 3 (A3). Key flow features: (1) the shock from the top wall deflection corner, (2) the leading edge separation shock, (3) the separating shear layer (4) the pressure wave of the present cycle. IP is the impinging point. The black square traces the pressure wave of the previous cycle. The green square mark the upstream traveling pressure wave of the present cycle.}
   \end{figure*}
     \begin{figure*}

	\includegraphics[scale=1.05]{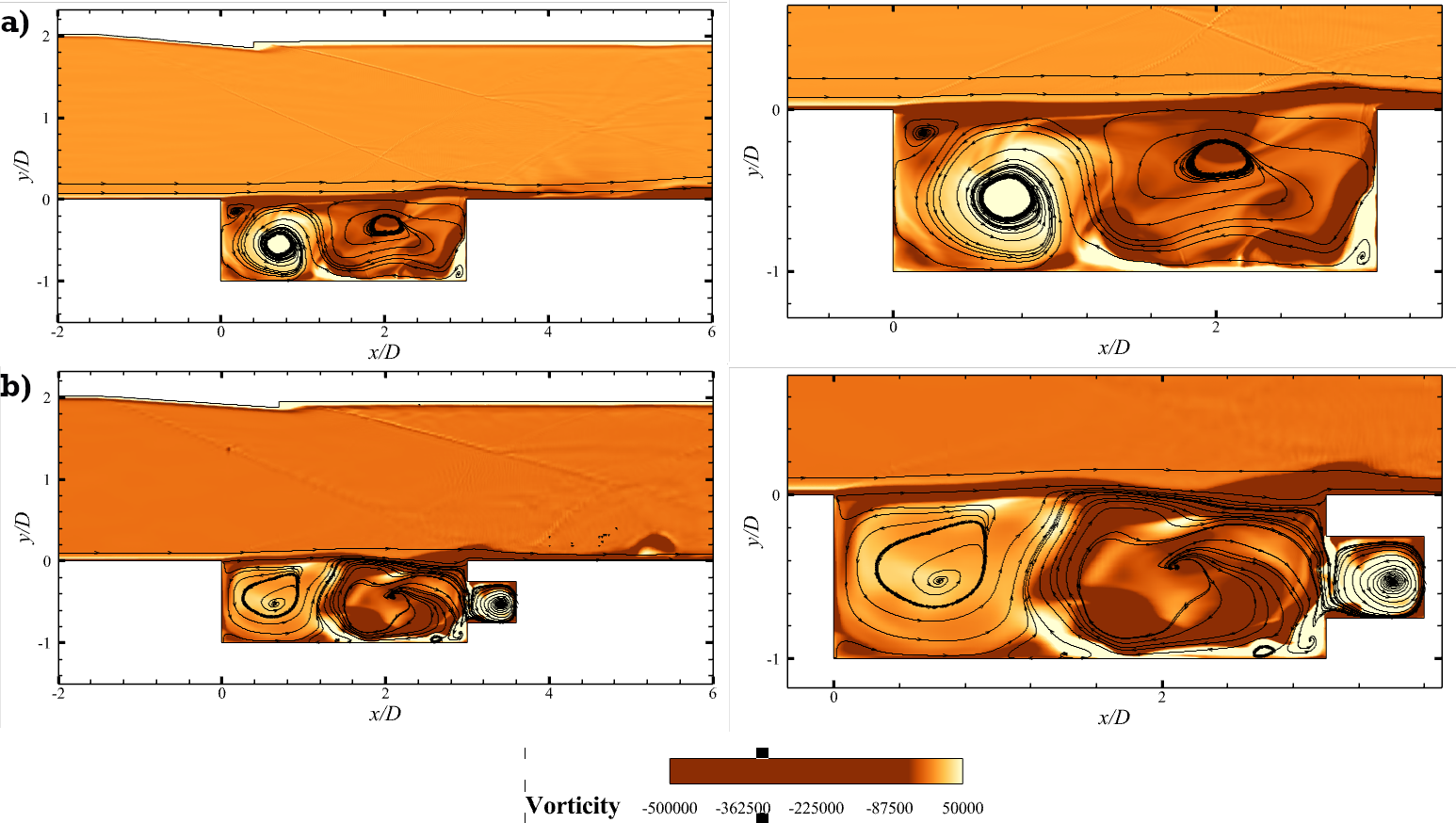}
    \centering

    \caption{\label{fig:29} Vorticity fields of the cavity a) without sub-cavity (R3) and b) with aft-wall sub-cavity (A3) at $M_\infty$=3. The close-up of the vorticity field with the streamlines near the aft edge of the cavity shows the difference in the flow field in the absence and presence of the sub-cavity at t/T= 118.81 and 128.04, respectively.}
   \end{figure*}  


\section{\label{sec:level3}Results and Discussion}

This section comprises four subsections, each addressing a key aspect of cavity flow physics for the given geometrical configurations at \(M_\infty = 2\) and \(3\). In Section~\ref{section:SA}, we perform spectral analysis using power spectral density and wavelet transforms to identify oscillation frequencies and compare the dominant frequency and its energy distribution across the reference and sub-cavity configurations. In Section~\ref{section:FV}, we visualize the flow field to examine how internal cavity flow responds to the presence of sub-cavities at both Mach numbers. Section~\ref{section:CA} presents cross-correlation analysis between pressure signals at probes P1 and P2, while Section~\ref{section:DMD} employs Dynamic Mode Decomposition as a reduced-order model. These analyses collectively reinforce the insights from the spectral and flow-field visualizations and help determine which sub-cavity configuration most effectively suppresses the dominant frequency at each \(M_\infty\).

\subsection{Spectral Analysis} \label{section:SA}
This section compares the frequency content of the reference cavity configuration (without sub-cavities) to that of configurations incorporating sub-cavities, for both \(M_\infty = 2\) and \(3\). This comparison enables the evaluation of the sub-cavities’ effectiveness in suppressing oscillation frequencies. We record pressure signals at each probe located within the cavity at time intervals ranging from \(3.5 \times 10^{-8}\) to \(4 \times 10^{-8}\) seconds, corresponding to a sampling rate of 25–30~MHz. This high sampling rate ensures wide frequency resolution and satisfies the Nyquist criterion. Each probe collects a sufficient number of samples to achieve a frequency resolution of 200~Hz, thereby capturing the dominant oscillation modes present in all configurations.

The Power Spectral Density (PSD) is calculated using the "pwelch" algorithm from the unsteady pressure data collected by probes P1, P2, and P3 for each configuration. The energy distribution in the unsteady pressure signal is analyzed using a Hanning window with 50\% overlap. We use the frequency (f) to the square of the freestream pressure ratio ($(p_\infty)^2$) to normalize the power spectral density (Gxx(p)). Additionally, for each scenario, we compare the Strouhal number (St)(fL/$U_\infty$) of the modes derived from the PSD analysis with the modified Rossiter's formula (equation \ref{eq:rossiter})\cite{rossiter1964wind}. For every configuration, the dominating Strouhal number and associated frequency are determined.

\begin{equation}
\frac{f L}{U_{\infty}} = \frac{n - \alpha}{\left( M_{\infty} \left( 1 + \frac{\gamma - 1}{2} M_{\infty}^2 \right)^{-0.5} + \frac{1}{\kappa} \right)}
\label{eq:rossiter}
\end{equation}

where, L is the characteristic length, $\kappa$=0.47 and $\alpha$=0.25 as suggested in the literature by Heller \cite{heller1971flow}.
 
 Continuous Wavelet Transformation (CWT) evaluation determines the temporal evolution of frequencies present in the system by providing a better time and frequency resolution. Chui \cite{Chui1992}, Torrence \cite{Torrence1998}, and Debnath \cite{Debnath2014} have elaborately studied the wavelet transformation and can be referred to for deeper insights. We normalize the frequency as Strouhal number (St)(fL/$U_\infty$) and time duration (t) with T (L/$U_\infty$).
\subsubsection{Power Spectral Density Analysis (PSD)}

Figs.~\ref{fig:4} and \ref{fig:5} present the power spectral density (PSD) obtained from probes P1, P2, and P3 for the reference configurations and the cases with sub-cavities located at the front wall and aft wall, for \(M_\infty = 2\) and \(3\), respectively. We retain these probe locations across all configurations, including those with sub-cavities, as their positions near the leading and trailing edges are critical to the flow dynamics. The shear layer originates from the leading edge and impinges on the trailing edge, forming a reattachment shock and enabling mass re-entry into the cavity. These interactions primarily govern the cavity’s oscillatory behavior. We compare the dominant frequency modes across configurations based on the PSD results, as summarized in Tables~\ref{tab:table3} and \ref{tab:table4}. For clarity, we denote each configuration based on its freestream Mach number and sub-cavity placement, as summarized in Table~\ref{tab:table1}. The reference case is labeled as \textbf{R}, the front-wall sub-cavity as \textbf{F}, and the aft-wall sub-cavity as \textbf{A}. Accordingly, configurations at \(M_\infty = 2\) are denoted as R2, F2, and A2, while those at \(M_\infty = 3\) are labeled R3, F3, and A3.

The key observations  for the $M_\infty$=2 are as follows :
\begin{itemize}

    \item Probe P2 consistently exhibits the highest energy content across each configuration, indicating stronger pressure oscillations near the aft cavity wall. The PSD values for Probe-P2 often reach up to 5-6 in the case of R2, reflecting significant acoustic activity.
 
    \item In the absence of sub-cavity (R2), the dominant St is at \(\mathrm{St2} = 0.55\). F2 exhibits amplified frequency of oscillations (\(\mathrm{St1} = 0.680\)).The PSD values, however, are comparatively lower, peaking near 0.04, indicative of energy redistribution over several higher modes. 
    
    \item The aft-wall sub-cavity configuration (A2) results in the most effective suppression of dominant oscillations at $M_\infty$=2. The dominant Strouhal number is reduced from  \(\mathrm{St1} = 0.55\) in R2 to \(\mathrm{St2} = 0.52\)  A2, which accounts for  a 5.45\% reduction. 
    
\end{itemize}

For $M_\infty=3$ cases, we derive the following observations from the PSD analysis:
\begin{itemize}
    \item Similar to $M_\infty=2$ cases, the probe P2  reports the maximum energy content and hence, the dominant mode for all the configurations. However, the PSD values here for the $M_\infty=3$ cases are lower than their  $M_\infty=2$ counterparts.
    \item The reference case R3 reports a higher dominant frequency mode than the $M_\infty=2$ cases. The higher freestream velocity accelerates the flow. The increase in $M_\infty$ enhances the compressibility effect, and also shifts the impinging shock downstream on the shear layer. Hence, the growth of the destabilizing KH instability in the shear layer is attenuated. Consequently, the increased speed and reduced enhancement of the KH instability cumulatively increase the frequency of oscillations \cite{bhaduri2025influence}.  
  
    \item For $M_\infty=3$,  the front-wall sub-cavity (F3) reduces the dominant St from \(\mathrm{St5} = 0.82\) in R3 to \(\mathrm{St1} = 0.63\) (a 23.15\% reduction). Although the sub-cavity at the aft-wall (A3) also reduces the oscillation frequency by  8.5\% (St4 = 0.75), the front-wall sub-cavity shows a more dominant suppressing effect for $M_\infty=3$.

\end{itemize}
The PSD analysis illustrates that for $M_\infty=2$, the aft-wall subcavity (A2) reduces the dominant frequency while the front-wall subcavity increases the dominant frequency. For $M_\infty=3$ , both the front-wall and aft-wall subcavities reduce the dominant frequency, but the front-wall attenuates it to a greater extent. We will further explore the energy distribution in the dominant frequency modes for all the cavity configurations.

\subsubsection{Continuous Wavelet Transform (CWT)}
In this section, we analyze the Continuous Wavelet Transform (CWT) results for the reference cases (R2 and R3) and the sub-cavity configurations at \(M_\infty = 2\) and \(3\). We place probes P1 and P2, along with probes i–vii, inside the cavity. Their locations, specific to each configuration, are provided in Table~\ref{tab:table2}. These probes record the temporal pressure fluctuations throughout the simulation. By examining the CWT of these signals, we capture the evolution of frequency within the cavity, identify the dominant frequency modes, and assess the energy content of these modes over the entire duration. The x-axis denotes time, while the y-axis represents the Strouhal number. The color scale of the spectrogram indicates the CWT coefficients, normalized by the maximum coefficient for each probe.

For \(M_\infty = 2\), all three cases, R2 (Fig.~\ref{fig:6}), F2 (Fig.~\ref{fig:7}), and A2 (Fig.~\ref{fig:8}) exhibit a dominant Strouhal number (\(\mathrm{St}\)) that remains consistent across most internal probes and agrees with the dominant frequency identified through PSD analysis (Fig.~\ref{fig:4}). We highlight this dominant \(\mathrm{St}\) using a white box in the spectrograms. Among the configurations, F2 shows the highest dominant Strouhal number of 0.680, followed by R2 at 0.550. A2 demonstrates a reduced dominant frequency at 0.540, indicating effective suppression. The colormap in each spectrogram represents the magnitude of the wavelet coefficients, which correspond to the instantaneous amplitude of oscillations. All cases display energetic high-frequency bands, representing intermittent secondary modes typically linked to turbulent flow structures. The global wavelet power spectra at probes P1 and P2 (Fig.~\ref{fig:9}) reveal that the reference case R2 exhibits stronger energy concentration at its dominant frequency compared to both sub-cavity cases. Despite F2 having a higher dominant frequency, its global wavelet power is lower, indicating weaker energy content in the dominant oscillation mode as well as in the overall frequency modes. This suggests that although the oscillation frequency increases, the oscillation strength or coherence does not.

The root mean square of acoustic impedance ($Z_\mathrm{rms}$), representing the local amplitude ratio of pressure to velocity fluctuations, serves as an indicator of the strength of acoustic wave interactions at specific locations. A higher $Z_\mathrm{rms}$ generally implies stronger pressure–velocity coupling, which can enhance the feedback loop within the cavity. Figure~\ref{fig:10}a shows that $Z_\mathrm{rms}$ is highest for the R2 case at both probe locations. At the aft-wall location (P2), the front-wall sub-cavity case (F2) exhibits a moderately higher $Z_\mathrm{rms}$ than A2, whereas at the front-wall location (P1), A2 shows higher impedance than F2.

To further interpret the acoustic behavior, we also evaluate the normalized acoustic impedance as:

\[
R_\mathrm{rms} = \mathrm{rms}\left( \frac{p'}{\rho_\infty c_\infty u'} \right),
\]

where $p'$ and $u'$ are the pressure and streamwise velocity fluctuations, respectively, $\rho_\infty$ is the freestream density, and $c_\infty$ is the speed of sound. This formulation captures the instantaneous, time-local coupling between pressure and velocity at each location. 
It quantifies how acoustically reflective or absorptive a region is relative to the freestream. A higher $R_\mathrm{rms}$ indicates stronger local coupling between pressure and velocity fluctuations, potentially supporting sustained oscillations. As shown in Figure~\ref{fig:10}b, $R_\mathrm{rms}$ is highest for A2 at P1 and for R2 at P2. In contrast, the F2 case exhibits a significant reduction in $R_\mathrm{rms}$ at both locations, indicating weaker local acoustic coupling.

For the $M_\infty = 3$ cases, the continuous wavelet spectrograms show that both F3 and A3 exhibit reduced dominant Strouhal numbers of 0.640 and 0.762, respectively, compared to 0.871 for the reference case R3 (Figs.~\ref{fig:11}–\ref{fig:13}). This trend aligns with the observations from the PSD analysis (Fig.~\ref{fig:5}). While a few internal probes exhibit slight variations in the dominant Strouhal number, the higher frequency bands appear less pronounced than in the $M_\infty = 2$ cases. 

The global wavelet power spectra (Fig.~\ref{fig:14}) reveal that R3 has the highest energy content at the dominant frequency near the front-edge (P1). In contrast, F3, despite having a lower dominant Strouhal number, shows relatively higher energy near the aft-edge (P2)(Fig.~\ref{fig:14}b).

At the front-wall location (P1), the acoustic impedance ($Z_\mathrm{rms}$) remains high for both R3 and A3 cases, as shown in Figure~\ref{fig:15}a. The R3 case also exhibits the highest normalized impedance ($R_\mathrm{rms}$), as shown in Figure~\ref{fig:15}b. At the aft-wall location (P2), the front-wall sub-cavity case (F3) shows a higher $Z_\mathrm{rms}$ compared to R3 and A3 (Figure~\ref{fig:15}a), while R3 again leads in $R_\mathrm{rms}$ (Figure~\ref{fig:15}b).

Notably, F3 shows a significantly lower $R_\mathrm{rms}$ at P1, and A3 has a similarly low $R_\mathrm{rms}$ at P2. These reductions indicate weaker local acoustic coupling at the respective edges, consistent with suppression of the feedback loop. Since cavity resonance at higher Mach numbers is primarily governed by compressibility-driven feedback mechanisms, the front-wall sub-cavity (F3) effectively disrupts the upstream-traveling acoustic waves at P1. This results in a stronger suppression of the dominant frequency, as observed in the spectral analysis.

Spectral analysis reveals the dominant frequency mode, its temporal evolution, and associated energy content for each configuration under investigation. The analysis also identifies the sub-cavity configurations that most effectively achieve passive control over cavity oscillation frequencies. The following section (Sec:\ref{section:FV}) presents numerical Schlieren visualizations to qualitatively examine differences in the flow fields that contribute to the alteration of the dominant frequency.
\subsection{Flow Visualisation} \label{section:FV}
The separating shear layer convects disturbances in the form of Kelvin-Helmholtz (KH) instability waves in open cavity flows ($L/D \le 10$). These waves are caused by shear, which destabilizes supersonic flow due to variations in flow characteristics between neighboring layers. The disturbances develop into a periodic array of compact spanwise vortices as they move in the direction of the trailing edge. Through nonlinear interactions, these vortices combine and disperse the vorticity field. A reattachment shock is produced, and the fluid is forced into the cavity when the disturbances strike the trailing edge. A pressure wave is created at the cavity floor close to the trailing edge by this mass entrainment. It spreads upstream within the cavity and causes the shear layer to develop ordered vortical patterns via acoustic-vortex resonance. This creates a feedback loop inside the cavity, which is crucial to the dynamics of the cavity flow and, consequently, to its use. These vortices improve fluid entrainment, speed up the spreading of the shear layer, and ease momentum and heat transfer as they combine or pair to create KH rolls downstream~\cite{heller1996cavity,krishnamurty1955acoustic,karthick2021shock,wang2013characteristics,thangamani2019mode,rockwell1978self}. The shock wave from the top wall interacts with the shear layer disturbances in confined supersonic cavities, changing the dynamics of the flow. Our earlier investigations ~\cite{bhaduri2024effects, bhaduri2025influence} show that the impinging shock enhances mixing and decreases the dominating frequency of the cavity oscillations by intensifying the KH instability within the shear layer.

Sub-cavities located along the cavity walls disrupt the feedback loop by modifying different components of the loop, depending on their placement. As the spectral analysis in Section~\ref{section:SA} demonstrated, these modifications shift the oscillation frequencies of the cavity. This section examines how the sub-cavities influence the internal flow fields for \(M_\infty = 2\) and \(M_\infty = 3\). As previously discussed, the periodic feedback loop remains a key feature across all configurations. We analyze the temporal variation of the normalized pressure \((p/p_\infty)\) recorded at probes P1 and P2 to identify a complete oscillation cycle. We then use normalized density gradient contours\(\left(\nabla \rho \cdot (D/\rho_{\infty})\right)\), corresponding to specific time instances within that cycle, to visualize the associated flow structures. We normalize time \((t)\) by \(T = L/U_\infty\), where \(L\) is the cavity length and \(U_\infty\) is the freestream velocity.

\subsubsection{Flow visualisation at $M_\infty$=2 }
Fig. \ref{fig:16} shows the temporal variation of the normalized pressure for the case R2. At the aft-wall (Fig.\ref {fig:16}b), the pressure is decreasing initially and reaching a minimum towards the end beyond (t/T=89), before increasing again. The front-wall pressure is more fluctuating , increasing and, at some instances, decreasing before it reaches a local maximum coincident with the minimum aft-wall pressure (Fig. \ref{fig:16}a). Beyond this, the front-wall pressure decreases again. Following the pressure variation near the front and aft-walls of the cavity, Fig. (\ref{fig:17}) shows the flow field of the cavity. At the beginning of the cycle considered, the shock (1) originating at the top wall deflection corner interacts with the separation shock(2) as the shear layer (3) separates from the leading edge. The compression wave then impinges on the shear layer at IP and reflects as an expansion wave. The subsequent compression and expansion of the flow enhances the KH instability, which is manifested as a very highly perturbed shear layer.  The pressure wave (4) from the previous cycle travels downstream, augmenting the perturbations in the shear layer. As the shear layer impinges on the aft-wall of the cavity, a reattachment shock (5) forms. The mass enters the cavity, leading to a pressure wave (6) at the trailing edge. This pressure wave (6) raises the pressure at the aft-wall at the beginning of this cycle, as seen in Fig. \ref{fig:16}b. Now, this pressure wave (6) travels upstream, as traced by the green square in the subFIG.s of Fig. \ref{fig:17}. It reaches the front-wall and reflects, between t/T =88.97 and 89.4 (Fig. \ref{fig:17}g- \ref{fig:17}i), perturbing the shear layer. The maximum pressure at the front-wall during this period is related to this incident. In Fig. \ref{fig:17}h, we can see the entrainment of mass inside the cavity resulting in a pressure wave (7) and initiating the next cycle. This pressure wave reflects increasing pressure at the aft-wall towards the end of this cycle (Fig. \ref{fig:16}b). The Strouhal number corresponding to the duration of this cycle is 0.554, which is nearly equal to the dominant one obtained from the spectral analysis. 

The introduction of the front-wall sub-cavity significantly alters the cavity flow behavior at $M_\infty$=2, where convective mechanisms dominate due to moderate compressibility. The pressure trace at probe P1 near the front edge (Fig.~\ref{fig:18}a) exhibits a more regular and lower-amplitude signal compared to the reference case, while probe P2 near the aft edge (Fig.~\ref{fig:18}b) shows reduced sharpness in peaks. These trends reflect the diminished strength of upstream-traveling pressure waves and suggest that the front-wall sub-cavity weakens the wave interaction with the separating shear layer. This suppression is further supported by the lower values of normalized impedance and the global wavelet power compared to the reference case, indicating that overall flow unsteadiness and turbulence intensity are reduced.  Fig. \ref{fig:19} shows the numerical schlieren for the entire cycle. The duration of this feedback cycle corresponds to St=0.681, which is nearly equal to that of the dominant St acquired from the spectral analysis. The flow visualization confirms that a portion of the upstream-traveling pressure wave is entrained into the sub-cavity. This interaction partially reflects and scatters the wave, thereby disrupting the coherence of vortex roll-up at an earlier phase. Compared to the reference case, the KH structures in the shear layer appear thinner and less perturbed, consistent with the reduced disturbance energy.  However, the dominant Strouhal number increases to $\mathrm{St} = 0.68$, higher than that in the reference case. The front-wall sub-cavity diverts a portion of the upstream-traveling wave, thereby weakening the pressure perturbations that interact with the separating shear layer. While this reduces the disturbance energy, it leads to the formation of smaller-scale Kelvin–Helmholtz (KH) structures that roll up earlier along the shear layer. These smaller vortices evolve and convect more rapidly across the cavity, effectively shortening the feedback loop duration. As a result, the dominant oscillation frequency increases, despite the overall weakening of the feedback strength. Thus, the front-wall sub-cavity acts as a wave-energy attenuator, dissipating part of the incoming wave before it re-initiates shear layer instability. Despite the elevated oscillation frequency, the flow field becomes less energetic, and the feedback loop is weakened in intensity, effectively suppressing the cavity’s self-sustained oscillations in terms of amplitude, though not in frequency.

The addition of the aft-wall sub-cavity modifies the cavity flow dynamics at $M_\infty = 2$. The pressure variation at probe P2 near the aft edge (Fig.~\ref{fig:20}b) confirms the presence of multiple characteristic waves. Fluid enters the cavity through the aft edge, and a portion of this mass enters the sub-cavity, generating local disturbances. The remaining portion contributes to the formation of the upstream-traveling pressure wave (5), whose path is traced by the green squares in the numerical schlieren (Fig.~\ref{fig:21}). This wave reaches the front wall before $t/T = 89$, as also evidenced by the pressure data at probe P1 near the front edge (Fig.~\ref{fig:20}a). Although an expansion wave (4) forms when the top-wall shock (1) impinges the shear layer (3) after interacting with the leading-edge separation shock (2) at the interaction point (IP), and the rolling structures in the shear layer indicate enhanced Kelvin–Helmholtz (KH) instability, the overall perturbation in the shear layer reduces in the presence of the aft-wall sub-cavity.

The sub-cavity entrains a part of the incoming mass and dissipates energy locally, thereby reducing the energy content of the pressure wave (5) that reaches and couples with the vortices in the shear layer. This attenuation weakens the feedback loop and suppresses oscillations. The duration of this cycle corresponds to $\mathrm{St}=0.521$, which matches the dominant frequency obtained from spectral analysis. Consistent with this suppression, the acoustic impedance analysis shows that the root mean square impedance ($Z_\mathrm{rms}$) for A2 is lower than the reference case (R2) at both P1 and P2 (Fig.~\ref{fig:10}a), indicating weaker pressure–velocity coupling at these locations. More notably, the normalized impedance ($R_\mathrm{rms} = Z_\mathrm{rms} / \rho_\infty c_\infty$) is significantly reduced at P2 for A2 (Fig.~\ref{fig:10}b), confirming diminished local acoustic reflectivity and, hence, a weakened feedback loop near the aft-wall.

At the lower Mach number ($M_\infty = 2$), the flow is predominantly governed by convective hydrodynamic mechanisms. The cavity oscillations in this regime are initiated by the roll-up of shear-layer vortices near the leading edge, which convect downstream and impinge on the aft wall, generating upstream-traveling pressure waves that close the feedback loop. Figure~\ref{fig:22} compares the vorticity fields of the reference (R2) and aft-wall sub-cavity (A2) cases at the beginning of their respective oscillation cycles. In the aft-wall sub-cavity configuration, the flow near the cavity ramp is significantly altered. A part of the vortex system is diverted into the sub-cavity, while the remaining flow contributes to the formation of vortices inside the cavity. The vortex inside the sub-cavity rotates in the opposite sense to the upstream-propagating vortical structures, leading to destructive interference. This scattering and redistribution of vorticity effectively weaken the generation of coherent upstream-traveling pressure waves, thereby attenuating the convective-acoustic coupling responsible for sustaining oscillations. As a result, the feedback loop is disrupted both by modifying the hydrodynamic impingement process and by reducing the acoustic response.
\subsubsection{Flow visualisation at $M_\infty$=3 }
For $M_\infty = 3$, the cavity configuration without a sub-cavity (R3) exhibits a strong pressure oscillation cycle. At the beginning of the cycle, a low-pressure region appears at the front wall (Fig.~\ref{fig:23}a), while the aft-wall shows a pressure peak (Fig.~\ref{fig:23}b). The front-wall pressure reaches a maximum at $t/T = 119.5$, whereas the aft-wall pressure is at a minimum. Shortly after, another peak appears at the aft wall between $t/T = 119.5$–120. The numerical schlieren (Fig.~\ref{fig:24}) shows that the top-wall shock (1) impinges the shear layer (3) after interacting with the leading-edge separation shock (2), with the interaction point (IP) located far downstream near $2.78D$. As reported in our earlier study~\cite{bhaduri2025influence}, this downstream impingement, coupled with strong compressibility effects, suppresses the spanwise growth of KH instabilities and limits amplification due to shock–shear interactions. The mass entrained into the cavity generates a pressure wave (5), seen as a peak at the aft wall, which propagates upstream (green squares, Fig.~\ref{fig:24}) and reaches the front wall to reinitiate the shear-layer disturbances. The full cycle duration corresponds to a Strouhal number of $\mathrm{St} = 0.826$, consistent with the dominant frequency from spectral analysis.

In the R3 case, both global wavelet spectra and impedance analysis support the presence of strong feedback. The global wavelet power at the dominant frequency is highest near the front wall (P1) (Fig.~\ref{fig:14}a), reflecting strong coupling at the initiation point of the feedback loop. Correspondingly, $Z_\mathrm{rms}$ remains high at both P1 and P2 (Fig.~\ref{fig:15}a), and $R_\mathrm{rms}$ is highest at P1 (Fig.~\ref{fig:15}b), confirming strong pressure–velocity coupling and high acoustic reflectivity, which reinforce the compressibility-driven feedback mechanism.

In the presence of the front-wall sub-cavity (F3), a distinct suppression mechanism emerges. At the start of the cycle (Fig.~\ref{fig:26}), the top-wall shock (1) impinges the shear layer (3) near the trailing edge (IP), while a pressure wave (4) from the previous cycle approaches the front edge, causing a pressure peak at P1 (Fig.~\ref{fig:25}a). As this wave reaches the front-wall, a portion of its energy enters the sub-cavity and reflects internally, while the remaining part interacts with the shear layer (Fig.~\ref{fig:26}e). This split in acoustic energy disrupts the compressibility-dominated feedback loop at its origin, weakening the pressure wave–vortex coupling. The re-emergence of the reflected wave at P1 produces two adjacent peaks near $t/T = 128$, evidencing this bifurcation in wave pathways. A new pressure wave (5), generated near the aft wall, travels upstream and again partially diverts into the sub-cavity (Figs. \ref{fig:26} g–i). This altered interaction results in a longer feedback cycle with a reduced frequency, corresponding to $\mathrm{St} = 0.632$.

The impedance metrics confirm this disruption. At P1, case F3 exhibits a significantly lower \( R_\mathrm{rms} \) (Fig. ~\ref{fig:15}b), indicating reduced acoustic reflectivity and a diminished potential to initiate shear-layer resonance. Although \( Z_\mathrm{rms} \) remains moderately high at P2 (Fig. ~\ref{fig:15}a), the weakened coupling at the front wall primarily governs the feedback dynamics. Therefore, the front-wall sub-cavity suppresses oscillations effectively by locally redirecting the incoming acoustic energy that would otherwise reinforce the feedback loop.

In contrast, the aft-wall sub-cavity configuration (A3) leads to a different suppression behavior. The pressure signal shows a reduced peak at the front wall during the cycle (Fig.~\ref{fig:27}a), and the aft-wall pressure falls as the upstream-traveling wave moves away from the trailing edge (Fig.~\ref{fig:27}b). Fig.~\ref{fig:28} illustrates the evolution of the density gradient magnitude from $t/T = 128.04$ to 129.37. At $t/T = 128.04$, the top-wall shock (1) impinges the shear layer (3) near the same downstream IP as in R3 and F3. However, mass entering through the aft wall is partially diverted into the sub-cavity and does not participate in the formation of the pressure wave. This flow diversion weakens the upstream-traveling wave and reduces shear-layer perturbations, leading to a lower-frequency feedback cycle, as seen in the vorticity fields in fig. \ref{fig:29}. Consistently, the acoustic impedance results show a marked drop in $R_\mathrm{rms}$ at P2 for A3 (Fig.~\ref{fig:15}b), confirming weaker pressure–velocity coupling and less efficient wave reflection at the aft wall.

These observations confirm that both sub-cavity configurations suppress the feedback loop at \( M_\infty = 3 \), but through fundamentally different mechanisms. In the aft-wall sub-cavity case (A3), part of the downstream convecting mass flux is diverted into the sub-cavity, which effectively reduces the energy available for generating strong upstream-traveling pressure waves. This diversion weakens the impingement process at the cavity ramp, leading to attenuation of the acoustic feedback. On the other hand, the front-wall sub-cavity (F3) interrupts the feedback loop closer to its source by partially scattering and redirecting the incoming pressure waves before they can amplify shear-layer instabilities. This action delays and weakens the acoustic excitation of the shear layer near the leading edge. The significantly reduced \( R_\mathrm{rms} \) at P1, the suppressed dominant frequency, and the altered wave paths observed in the flow visualizations support this interpretation. Under highly compressible flow conditions, the front-wall sub-cavity (F3) is particularly effective because it intervenes at the point where the compressibility-driven acoustic modes interact with the shear layer \cite{sarkar1995stabilizing,Sandham_Reynolds_1991,tam1978tones,papamoschou1988compressible}, while the aft-wall sub-cavity (A3) primarily influences the hydrodynamic side of the loop by altering vortex impingement and momentum transfer.
\begin{figure*}
\centering
	\includegraphics[scale=1.3]{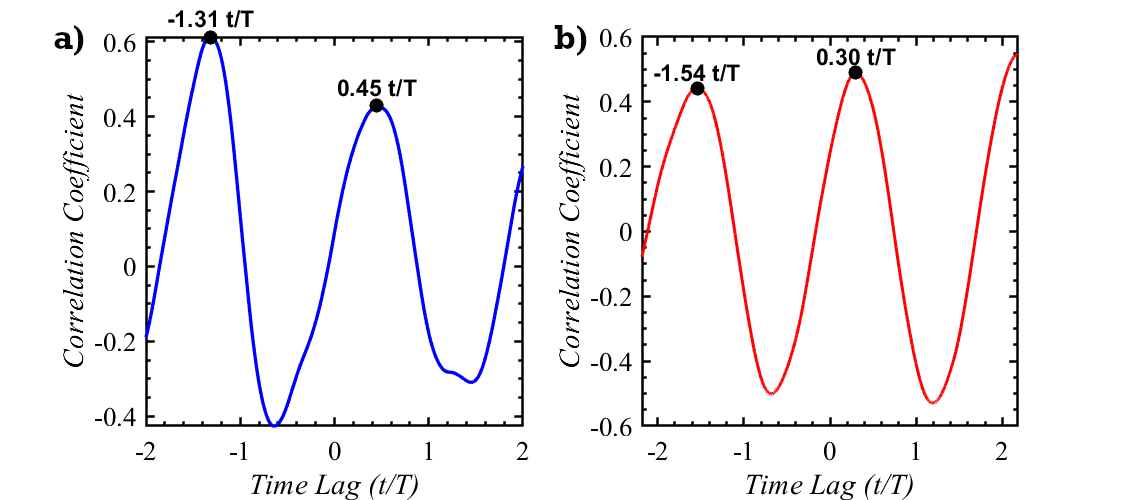}

    \caption{\label{fig:30}Cross correlation between the probes P2 and P1 for cavity  a) without sub-cavity (R2) and b) with aft-wall sub-cavity (A2) at $M_\infty$=2. }
   \end{figure*}
     \begin{figure*}

	\includegraphics[scale=1.3]{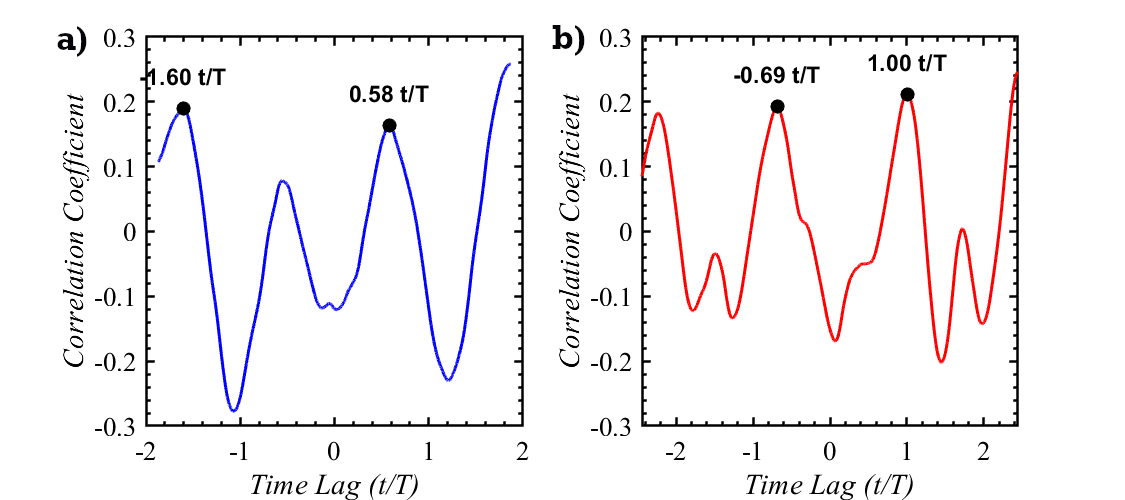}
    \centering

    \caption{\label{fig:31} Cross correlation between the probes P2 and P1 for cavity a) without sub-cavity (R3) and b) with front-wall sub-cavity (F3) at $M_\infty$=3.}
   \end{figure*}
   \begin{figure*}
  \centering
	\includegraphics[scale=0.85]{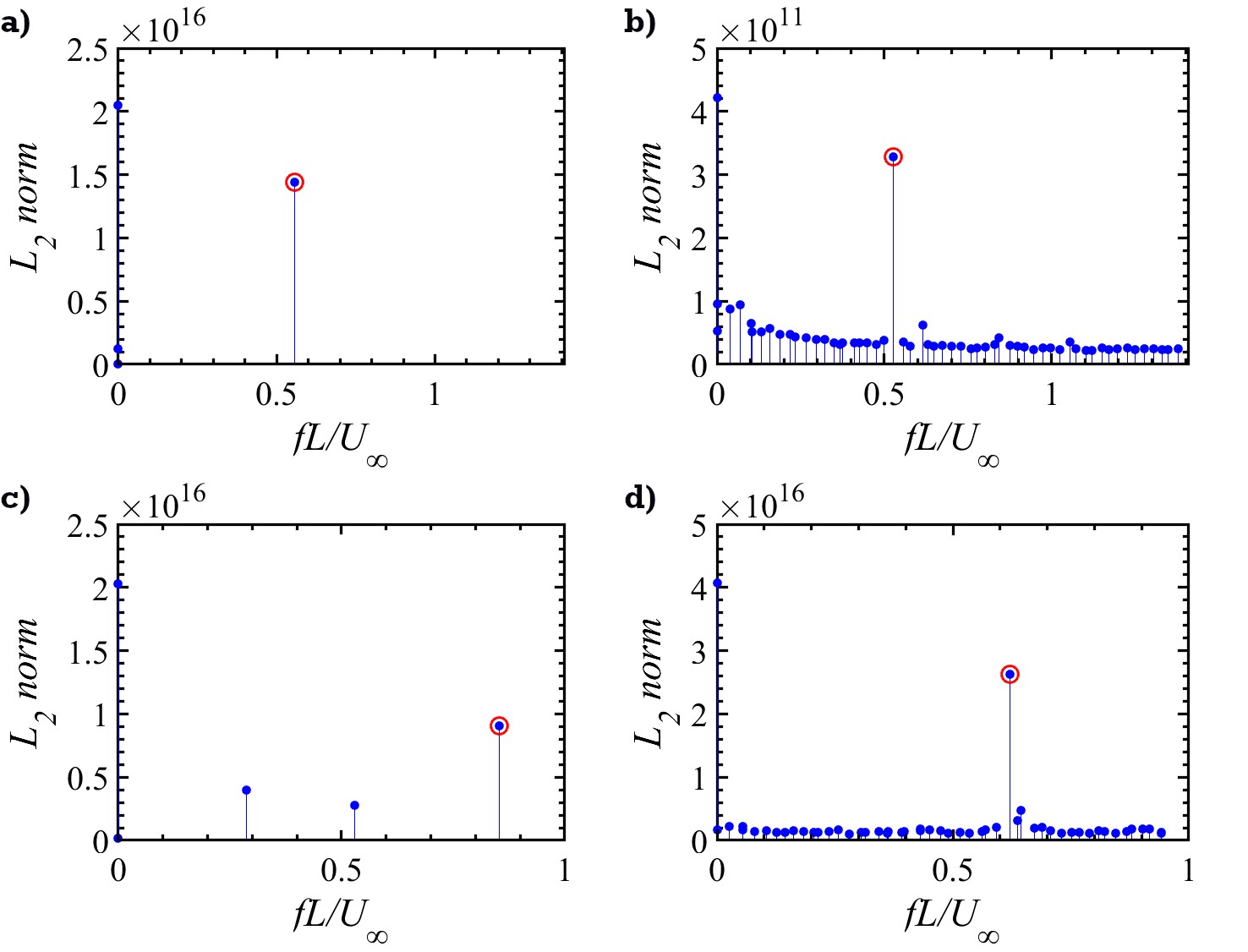}

    \caption{\label{fig:32} $L_2$ norm for cavity configurations a) R2 b) A2 c) R3 and d) F3, where the dominant frequency is marked.}
   \end{figure*}
   \begin{figure*}
\centering

	\includegraphics[scale=0.85]{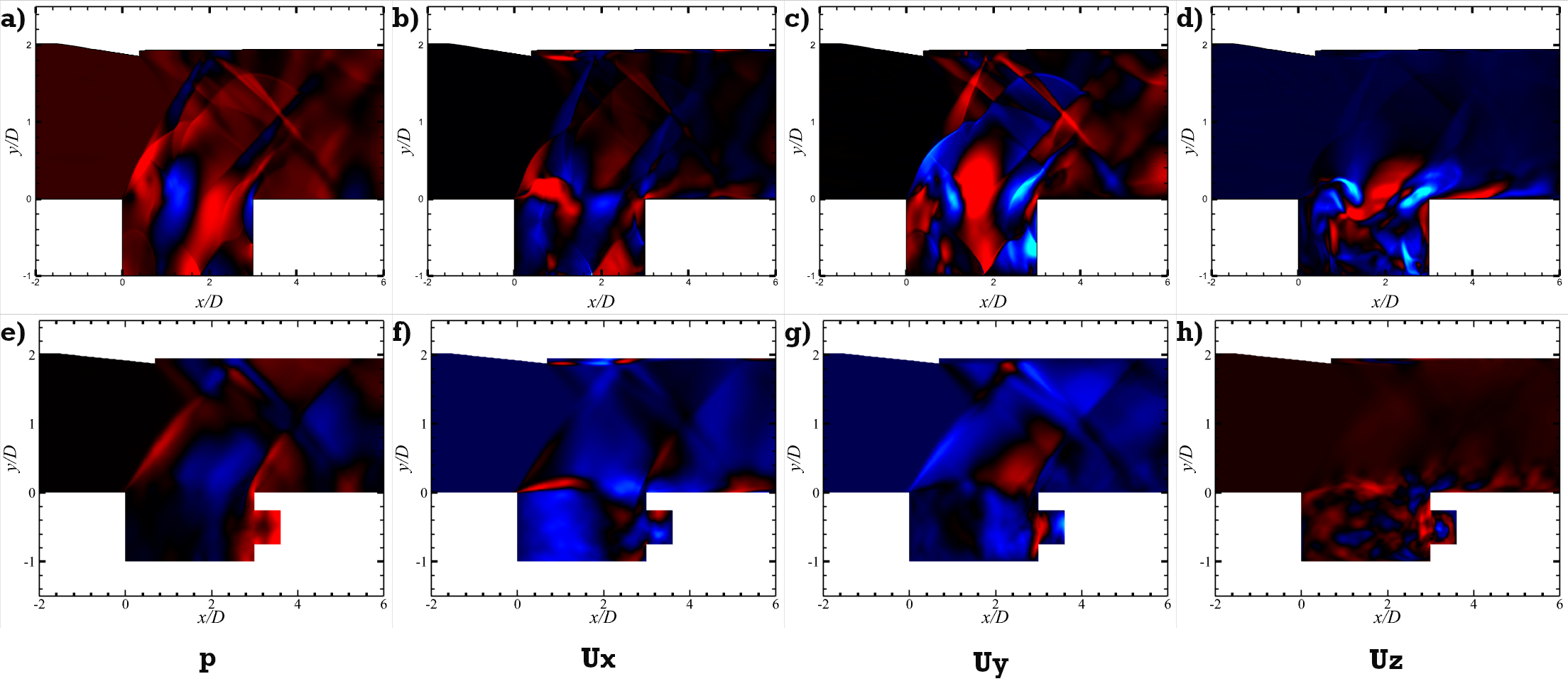}
    
    \caption{\label{fig:33} Pressure and velocity fields corresponding to the dynamic mode related to the dominant mode for cavity  (a)-(d) without sub-cavity (R2)  and (e)-(h)  with aft-wall sub-cavity (A2) at M$_\infty$ = 2.}
   \end{figure*}
  \begin{figure*}
\centering

	\includegraphics[scale=0.85]{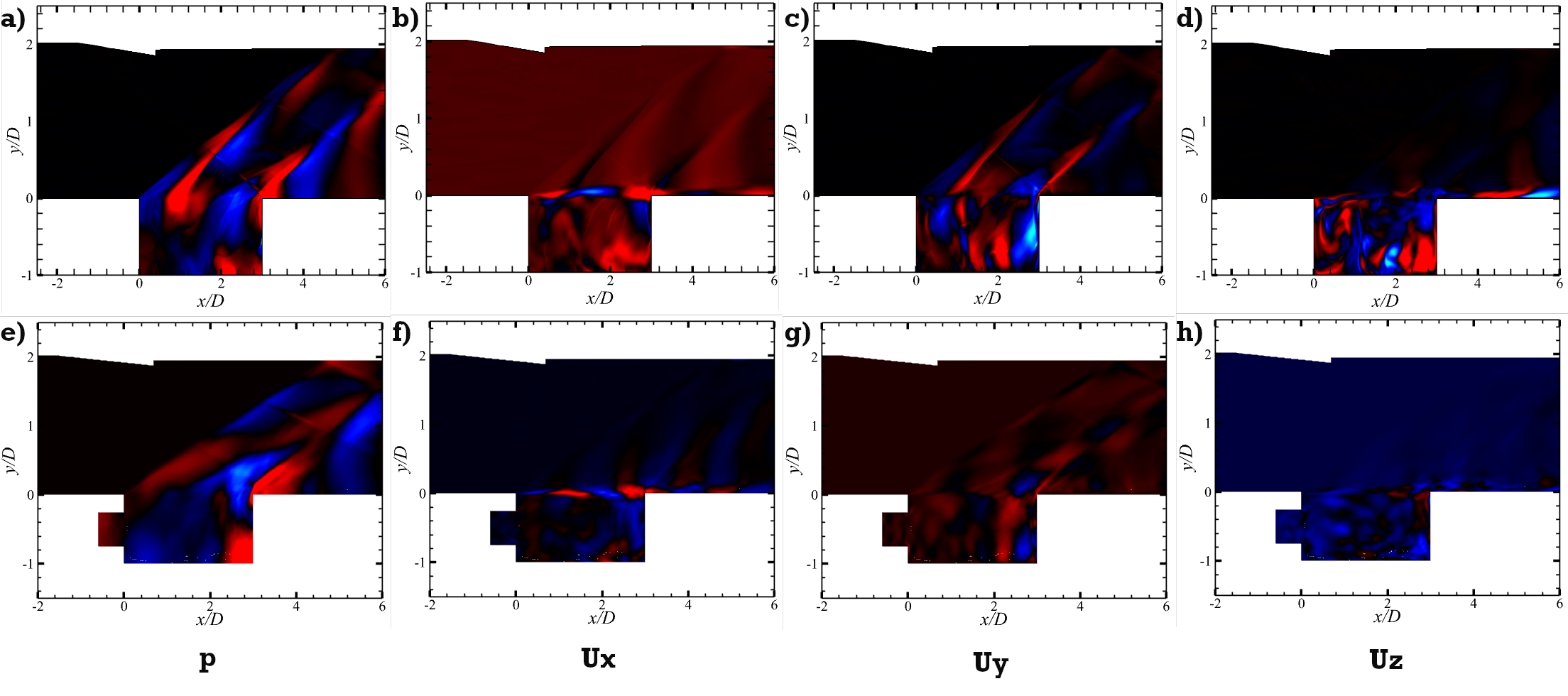}
    
    \caption{\label{fig:34} Pressure and velocity fields corresponding to the dynamic mode related to the dominant mode for cavity  (a)-(d) without sub-cavity (R3)  and (e)-(h)  with front-wall sub-cavity (F3) at M$_\infty$ = 3.}
   \end{figure*}


The numerical schlieren images and pressure signals at probe P1 (front wall) and probe P2 (aft wall) highlight how the sub-cavities alter the flow field and modulate the oscillation dynamics. The effectiveness of each control strategy depends on the nature of the feedback loop, which shifts with \( M_\infty \). At \( M_\infty = 2 \), the feedback is predominantly governed by convective hydrodynamic mechanisms namely, the downstream convection of shear-layer vortices and their role in initiating upstream-propagating acoustic waves. In this regime, the aft-wall sub-cavity more effectively suppresses the dominant oscillation by disrupting vortex impingement and altering the convective return path. In contrast, at \( M_\infty = 3 \), compressibility effects become dominant, and the feedback loop is primarily driven by direct acoustic interactions \cite{sarkar1995stabilizing,Sandham_Reynolds_1991,tam1978tones,papamoschou1988compressible}. Here, the front-wall sub-cavity achieves greater suppression of the oscillation by intercepting and scattering the incident pressure waves, thereby weakening their coupling with the shear layer and delaying disturbance amplification at the leading edge.
Further, we perform the cross-correlation and dynamic mode decomposition for the reference cases R2 and R3 and compare them with the aft-wall subcavity configuration for $M_\infty=2$ (A2) and front-wall subcavity configuration for $M_\infty=3$ (F3), respectively.
\subsection{Correlation analysis}\label{section:CA}
This section explores the cross-correlation analysis between probes P2 and P1 for the reference cases with the sub-cavity configuration having the least dominant Strouhal number at the respective $M_\infty$, to gain a better understanding of the feedback loop. Due to their advantageous positions near the leading and trailing edges, respectively, these probes are frequently chosen in all configurations, including sub-cavity scenarios. Shear layer impingement and reattachment shock creation cause flow to re-enter the cavity close to the trailing edge, whereas the shear layer itself originates at the leading edge. These two areas are essential for maintaining the feedback-driven cavity oscillations, making them crucial for describing the dynamics of the flow.

Figure~\ref{fig:30} shows the cross-correlation between the pressure signals at the aft wall (P2) and the front wall (P1) for the reference (R2) and aft-wall subcavity (A2) cases at \( M_\infty = 2 \). Two dominant peaks are identified in each case, corresponding to the downstream convective time lag and the upstream acoustic return time. In the reference configuration (R2), the maximum correlation occurs at a negative lag of \(-1.31~t/T\), indicating that the front-wall signal (P1) leads the aft-wall signal (P2). This behavior reflects a conventional feedback loop, where shear-layer vortices shed from the cavity's leading edge convect downstream, impinge upon the aft wall, and generate upstream-traveling acoustic waves. A secondary correlation peak is observed at a positive lag of \(+0.45~t/T\), representing the acoustic return time from the aft wall back to the front wall.In the aft-wall subcavity configuration (A2), the correlation pattern undergoes significant changes. The downstream hydrodynamic lag becomes more negative, shifting to \(-1.54~t/T\), indicating slower  convective transport of shear-layer vortices toward the aft wall. This suggests that the subcavity at the aft wall disrupts vortex impingement, possibly due to local vortex scattering near the subcavity. Additionally, the positive lag reduces to \(+0.30~t/T\), indicating a faster transport of the pressure wave to the front wall. This change is attributed to the subcavity’s influence on the impingement process, which modifies the generation and upstream propagation of the feedback wave potentially leading to a weaker or more diffused acoustic wave

These trends are consistent with the frequency suppression observed in the spectral data, confirming that the aft-wall subcavity effectively disrupts the loop by altering both the hydrodynamic and acoustic paths.The results highlight that at \(M_\infty = 2\), the feedback loop remains primarily convective, with control mechanisms having a significant influence on the vortex dynamics.


Figure~\ref{fig:31} shows the cross-correlation between the pressure signals at the aft wall (P2) and the front wall (P1) for the reference (R3) and aft-wall subcavity (F3) cases at \( M_\infty = 3\). Similar to the \(M=2\) case, two dominant peaks are identified, but the lag behavior shows the opposite trend.In the reference configuration (R3), the maximum correlation occurs at a positive lag of \(+0.58~t/T\), suggesting that the aft-wall signal (P2) leads the front-wall signal (P1). This pattern is typical of a feedback loop dominated by upstream acoustic propagation, where pressure waves generated at the aft wall travel back to the leading edge. In the front-wall subcavity case (F3), the positive lag increases to \(+1.00~t/T\), indicating a delayed acoustic interaction at the front wall. This is consistent with the observation that the incident acoustic energy is diverted into the front-wall subcavity, delaying the onset of shear layer disturbances that contribute to the feedback loop. Simultaneously, the negative lag shifts from \(-1.60~t/T\) in the reference case to \(-0.69~t/T\) in the front-wall controlled case, implying a faster or less coherent hydrodynamic disturbance propagation. The shift in lag behavior at \(M_\infty=3\) suggests that under highly compressible conditions, the feedback loop becomes more acoustically dominated. In this regime, the front-wall subcavity primarily affects the acoustic initiation process, leading to frequency suppression by delaying the upstream wave interaction with the shear layer.

\subsection{Dynamic Mode Decomposition} \label{section:DMD}
In the previous sections, we quantitatively identified the most effective passive control strategies for the confined cavity at $M_\infty= 2$ and 3 
 using power spectral density analysis. The energy distribution and temporal evolution of the dominant modes for each configuration were subsequently examined through continuous wavelet analysis (Section \ref{section:SA}). Flow visualizations (Section \ref{section:FV}) revealed how the introduction of sub-cavities altered the flow field. Cross-correlation between signals from probes located near the front and aft edges of the cavity confirmed the efficient suppression of the feedback loop frequency caused by the aft-wall sub-cavity for $M_\infty=2$, and by the front-wall sub-cavity for $M_\infty=3$ (Section \ref{section:CA}). In order to determine the spatiotemporal coherence of the flow structures at the dominant frequency, we additionally perform the Dynamic Mode Decomposition (DMD) analysis on the pressure and velocity fields derived from LES simulations.

We have used the methods of snapshots where the snapshots are arranged in matrices and converted to an orthogonal and an upper triangular matrix (QR decomposition) followed by a Singular Value Decomposition (SVD) to extract the dynamic modes present in the system \cite{sayadi2016parallel,soni2019modal,bhatia2019numerical,sharma2024investigation,arya2023acoustic,kutz2016,brunton2016}.  The snapshots of pressure and velocity fields from the LES simulations are recorded at uniform time intervals of \(2.5 \times 10^{-5}\,\text{s}\), corresponding to a sampling frequency of \(F_s = 40\,\text{kHz}\). According to the Nyquist criterion, this allows the capture of frequency content up to \(F_s/2 = 20\,\text{kHz}\), which defines the upper limit of the resolved spectrum. A total of 200 snapshots are used, resulting in a total sampling duration of \(T = 200 \times 2.5 \times 10^{-5} = 0.005\,\text{s}\). The resulting frequency resolution is therefore \(\Delta f = 1/T = 200\,\text{Hz}\), allowing the DMD analysis to resolve spectral components in the range of 200 Hz to 20 kHz. These snapshots are organized into a data matrix, on which Singular Value Decomposition (SVD) is performed to extract the dominant spatiotemporal modes via the Dynamic Mode Decomposition (DMD) algorithm.
 The imaginary parts of the eigenvectors represent the frequency content of each mode, while the real parts indicate the modal behavior. The eigenvalues, interpreted through the \(L_2\) norm, help to identify the most energetic frequency modes.
The dominant frequency mode associated with the feedback mechanism governing cavity oscillations is identified based on the \(L_2\) norm of the DMD modes, which reflects the relative amplitude of each mode. The frequency corresponding to the highest \(L_2\) norm aligns well with the dominant peak observed in the PSD analysis. Figure~\ref{fig:32} presents the \(L_2\) norm distributions for all cavity configurations considered at \(M_\infty = 2\) and \(M_\infty = 3\). The most prominent modes appear at \(St = 0.55\) for R2, \(St = 0.528\) for A2, \(St = 0.857\) for R3, and \(St = 0.62\) for F3, closely corresponding to the spectral peaks observed in the PSD.

Fig. \ref{fig:33} presents the dominant dynamic mode extracted using DMD for the cavity configuration at \(M_\infty = 2\), without (R2, top row) and with (A2, bottom row) the aft-wall sub-cavity. Subfigures \ref{fig:33}a–d correspond to the instantaneous spatial fields of pressure, streamwise velocity ($U_x$), lateral velocity ($U_y$), and spanwise velocity ($U_z$) fields for the R2 case. The fields are not normalized.  The leading-edge shock, the top-wall deflection-induced shock, the expansion wave close to the impingement point, and the reattachment shock at the aft-wall interact to produce the alternating zones of compression and expansion inside the cavity that are captured by the pressure mode. The convective motion of large-scale vortices moving downstream along the shear layer and recirculating inside the cavity is highlighted by the $U_x$ mode. The $U_y$ structures appear anti-symmetric to the streamwise ones, characteristic of shear-layer roll-up. The spanwise undulations seen in the $U_z$ mode are linked to Kelvin-Helmholtz (KH) rolls and are exacerbated by compressive-expansive dynamics brought on by the impinging shock.

The comparable dynamic modes for the A2 configuration, which include the aft-wall sub-cavity, are displayed in the second row (Figs.\ref{fig:33}e–\ref{fig:33}h). These modes show lower intensity and larger structures than the R2 example. This results from modifications to the flow dynamics brought about by the aft-wall sub-cavity. The incoming flow splits at the sub-cavity entrance, according to flow visualizations: part of the flow enters the sub-cavity and forms a vortex that rotates in the opposite direction of the upstream-traveling feedback vortices. This counter-rotating vortex weakens the feedback loop by destructively interfering with it. As a result, the dominant DMD modes exhibit lower energy and less spatial complexity, which is consistent with the suppression of oscillations caused by feedback. These findings are consistent with the cross-correlation analysis's patterns.

The first row (Figs.\ref{fig:34}a–\ref{fig:34}d) presents the dominant dynamic modes for the R3 configuration (reference case without passive control), while the second row (Figs.\ref{fig:34}e–\ref{fig:34}h) shows the corresponding modes for the F3 configuration, which includes a front-wall sub-cavity at $M_\infty$=3.In the R3 scenario, the cavity and shear layer region exhibit strong modal activity for all variables, including the instantaneous spatial fields of pressure (p) and velocity components ($U_x$,$U_y$,$U_z$). High-amplitude structures close to the aft wall are especially highlighted by the pressure and streamwise velocity components (Figs. ~\ref{fig:34} a and b), indicating an active feedback mechanism. These structures exhibit self-sustained oscillations that are typical of open cavity flows at high Mach numbers. They are compact, propagate upstream, and span a variety of wavelengths.
In contrast, the F3 case with the front-wall sub-cavity shows even weaker mode amplitudes across all flow variables (Figs. ~\ref{fig:34}e–~\ref{fig:34}h). The pressure field and streamwise velocity reveal diminished high-energy features, while vertical and spanwise velocity components ($U_y$, $U_z$) exhibit notably less spatial structure. The front-wall sub-cavity scatters and partially absorbs the acoustical energy of upstream-traveling pressure waves before they can couple with the shear layer. This interferes with the receptivity of the shear layer near the front edge and suppresses the frequency of the oscillations. The conclusion that the front-wall sub-cavity interferes with the compressibility-driven feedback process at $M_\infty=3$ is supported by these DMD results, which are in line with the longer time delays and less coherence observed in the cross-correlation study for the F3 case.
\FloatBarrier
\section{\label{sec:level4}Conclusion}
In the present study, we investigate the passive control of oscillation frequencies in a confined supersonic cavity with a length-to-depth ratio of \( L/D = 3 \) at freestream Mach numbers \( M_\infty = 2 \) and \( 3 \), by incorporating sub-cavities with a length ratio \( l/L = 0.2 \). The analysis begins with the power spectral density (PSD) of unsteady pressure signals recorded at probes P1 and P2, positioned near the front and aft edges of the cavity, respectively. At \( M_\infty = 2 \), the aft-wall sub-cavity configuration (A2) effectively suppresses the dominant oscillation frequency, whereas the front-wall configuration (F2) results in frequency amplification. However, global wavelet power spectra reveal a lower amplitude at the dominant Strouhal number in the F2 case, and the normalized impedance values at both P1 and P2 are reduced, indicating weaker local acoustic coupling.

Flow visualizations using numerical Schlieren and vorticity fields reveal that, for \( M_\infty = 2 \), the feedback loop is primarily driven by vortices generated near the aft edge due to mass entrainment from the external flow. The A2 configuration intercepts part of this entrained flow, thereby diminishing its contribution to the feedback loop. Furthermore, the vorticity induced within the sub-cavity rotates in a direction opposite to that of the dominant cavity vortices, counteracting their strength and thus attenuating the upstream-propagating pressure waves. This mechanism weakens the feedback loop and leads to a reduction in the oscillation frequency. In contrast, while the F2 configuration partially blocks the acoustic energy transfer from the pressure wave to the shear layer, resulting in lower pressure amplitudes, it also accelerates the feedback cycle, leading to a higher dominant frequency.

At the higher Mach number \( M_\infty = 3 \), both the front-wall (F3) and aft-wall (A3) sub-cavity configurations result in a reduction of the dominant frequency. Numerical Schlieren fields show that in the F3 case, a portion of the upstream-traveling pressure wave is deflected into the sub-cavity, effectively reducing the acoustic energy available for shear-layer excitation. This results in weaker hydrodynamic–acoustic coupling and attenuates the dominant oscillation frequency characteristic of compressibility-driven feedback loops. Although the A3 configuration similarly disrupts the feedback by diverting some of the entrained mass in the aft wall sub-cavity, the F3 configuration proves more effective at \( M_\infty = 3 \), where compressibility effects dominate the feedback mechanism. Thus, suppressing upstream-propagating acoustic waves central to feedback dynamics in high-Mach flows is more effectively accomplished by modifying the interaction at the cavity’s front edge.

Cross-correlation analysis between probes P1 and P2 reveals increased time delays and enhanced coherence in cases A2 and F3, indicating modifications to the feedback loop dynamics. As established in the preceding spectral and flow-field analyses, the dominant frequency arises from the self-sustained feedback mechanism. The suppression of high-energy coherent structures in the dynamic modes of pressure and velocity fields extracted using Dynamic Mode Decomposition (DMD) corroborates the disruption of this feedback. These findings confirm the efficacy of the A2 and F3 sub-cavity configurations in attenuating oscillation frequencies by weakening the underlying feedback mechanism.

Future studies will examine the influence of the top-wall deflection angle, which remains fixed at $2.29^\circ$ in the present work.

\begin{acknowledgments}
The authors acknowledge the National Supercomputing Mission (NSM) for computational resources on 'PARAM Sanganak' at IIT Kanpur, supported by MeitY and DST, Government of India. They also thank the IIT Kanpur Computer Center for additional resources.
\end{acknowledgments}

\section*{Conflict of Interest}
The authors have no conflict of interest to disclose.

\section*{Data Availability}
The data that support the findings of this study are available from the corresponding author upon reasonable request.
\vspace{0.8em}

\noindent\textbf{Nomenclature} 

\vspace{0.5em}

\begin{table}[H] 
\noindent
\begin{tabular}{@{}ll@{}} 
$D$ & Depth of the cavity \\
$L$ & Length of the cavity \\
$l$  & Length of the sub-cavity \\
$\beta$ & Shock angle \\
$\theta$ & Deflection angle \\
$H_i$ & Height of the inlet \\
$H_o$ & Height of the outlet \\
$p_\infty$ & Freestream pressure \\
$U_\infty$ & Freestream velocity \\
$M_\infty$ & Freestream Mach number \\
$\langle p \rangle$ & Average pressure \\
$s$ & Radial distance \\
$\rho_\infty$ & Freestream density \\
$\nabla\rho$ & Gradient of density \\
$t$ & Time \\
$T$ & Reference time \\
$f$ & Frequency \\
$\dfrac{fL}{U_\infty}$ & Strouhal number (St) \\
$G_{xx}(p)$ & Power Spectral Density \\
\end{tabular}
\end{table}

\FloatBarrier  

\section*{References}

\nocite{*}
\bibliography{aipsamp}

\end{document}